\documentclass[aps,prd,twocolumn,nofootinbib,superscriptaddress,10pt]{revtex4-1}
\usepackage{amstext}
\usepackage{amssymb}
\usepackage{amsmath}
\usepackage{graphicx}
\usepackage{array}
\usepackage{subfigure}
\usepackage{color}
\usepackage{listings}
\usepackage{natbib}
\usepackage{url}
\usepackage{bm}
\usepackage{multirow}
\usepackage{etoolbox}
\usepackage[utf8]{inputenc}
\usepackage[colorlinks=true]{hyperref}
\usepackage[nolist]{acronym}

\usepackage{ulem}
\newcommand{\mtot}{M_\mathrm{total}}
\newcommand{\be}{\begin{equation}}
\newcommand{\ee}{\end{equation}}
\newcommand{\bea}{\begin{eqnarray}}
\newcommand{\eea}{\end{eqnarray}}
\def\cross{\times}

\begin{document}

\title{Searching for gravitational waves from compact binaries with
  precessing spins}

\author{Ian~Harry} 
\email{ian.harry@aei.mpg.de}
\affiliation{Max Planck Institute for Gravitational Physics (Albert Einstein Institute),
  Am M\"uhlenberg 1, D-14476 Potsdam-Golm, Germany}
  
\author{Stephen~Privitera} 
\email{stephen.privitera@aei.mpg.de}
\affiliation{Max Planck Institute for Gravitational Physics (Albert Einstein Institute),
  Am M\"uhlenberg 1, D-14476 Potsdam-Golm, Germany}

\author{Alejandro~Boh\'e} 
\email{alejandro.bohe@aei.mpg.de}
\affiliation{Max Planck Institute for Gravitational Physics (Albert Einstein Institute),
  Am M\"uhlenberg 1, D-14476 Potsdam-Golm, Germany}

\author{Alessandra~Buonanno} 
\email{alessandra.buonanno@aei.mpg.de}
\affiliation{Max Planck Institute for Gravitational Physics (Albert Einstein Institute),
  Am M\"uhlenberg 1, D-14476 Potsdam-Golm, Germany}


\begin{abstract}
Current searches for gravitational waves from compact-object binaries
with the LIGO and Virgo observatories employ waveform models with spins
aligned (or antialigned) with the orbital angular momentum.  Here, we
derive a new statistic to search for compact objects carrying generic
(precessing) spins. Applying this statistic, we construct banks of
both aligned- and generic-spin templates for binary black holes and
neutron star--black hole binaries, and compare the effectualness of
these banks towards simulated populations of generic-spin systems. We
then use these banks in a pipeline analysis of Gaussian noise to
measure the increase in background incurred by using generic- instead
of aligned-spin banks. Although the generic-spin banks have roughly a factor
of ten more templates than the aligned-spin banks, we find
an overall improvement in signal recovery at a fixed false-alarm rate
for systems with high-mass ratio and highly precessing spins.
This gain in sensitivity
comes at a small loss of sensitivity ($\lesssim$4\%) for systems that
are already well covered by aligned-spin templates. Since the
observation of even a single binary merger with misaligned spins could
provide unique astrophysical insights into the formation of these
sources, we recommend that the method described here be developed
further to mount a viable search for generic-spin binary mergers in
LIGO/Virgo data.
\end{abstract}

\maketitle


\acrodef{SMS}[sky-maxed SNR]{sky-maximized signal-to-noise ratio}
\acrodef{CBC}[CBC]{compact binary coalescence}
\acrodef{GW}[GW]{gravitational-wave}
\acrodef{PSD}[PSD]{power spectral density}
\acrodef{BNS}{binary neutron star}
\acrodef{NSBH}{neutron star-black hole}
\acrodef{BBH}{binary black hole}
\acrodef{PN}{post-Newtonian}
\acrodef{SNR}{signal-to-noise ratio}
\acrodef{FFT}{fast Fourier transform}
\acrodef{LIGO}{Laser Interferometer Gravitational-wave Observatory}
\acrodef{PDF}{probability density function}

\section{Introduction}
\label{sec:intro}

On September 14, 2015, the \ac{LIGO}~\cite{Harry:2010zz,
  TheLIGOScientific:2014jea} made its first observation of
gravitational waves, which were emitted by the binary black-hole
merger dubbed GW150914~\cite{Abbott:2016blz}. In the coming years, the
field of \ac{GW} astronomy will begin to take off in earnest, with
further upgrades to the LIGO detectors under way, and the expansion of
the \ac{GW} observatory network to include Advanced Virgo~\cite{TheVirgo:2014hva}, KAGRA~\cite{Somiya:2011np, Aso:2013eba} and
an additional LIGO observatory in India~\cite{LigoIndia}. We expect not only to
observe more binary black-hole mergers~\cite{Abbott:2016nhf}, but also
signals from binary neutron-star and neutron star--black hole
mergers~\cite{Abadie:2010cf}.

Compact-object binary systems (henceforth, compact binaries) are
thought to form through two channels: the coevolution of two massive stars in a
binary~\cite{Belczynski:2010tb, Dominik:2012kk, Kinugawa:2014zha,
  Ferdman:2014rna,Marchant:2016wow}, or by the dynamical capture of
two preformed compact objects in dense stellar environments such as
globular clusters~\cite{Rodriguez:2015oxa, Belczynski:2014iua,
  Clausen:2012zu, Ivanova:2007bu, Pooley:2003zb}. The relative and
absolute rates for these two potential formation channels are highly
uncertain; they depend sensitively on a number of poorly constrained
parameters, such as the typical stellar metallicity at formation, the
distribution of supernova kicks, and the binding energy of the common
envelope.  Indeed, clarification of the astrophysics of these
formation channels is one of the great scientific promises of
observing \ac{GW} signals from compact-binary
mergers~\cite{Belczynski:2015tba, Vitale:2015tea,
  Stevenson:2015bqa,TheLIGOScientific:2016htt}.  The parameters of the merging
binary are measurable through an observed \ac{GW}
signal~\cite{TheLIGOScientific:2016wfe,Veitch:2014wba}, which can then provide
information about the formation processes of the system.

In particular, it is thought that compact binaries formed by dynamical
capture are more likely to have component angular momenta (spins) at
large angles to the orbital angular momentum, while those formed by
common evolution are more likely to have spins that are nearly aligned
with the orbital angular
momentum~\cite{Belczynski:2007xg,Steiner:2011vr,Fragos:2010tm,Pooley:2003zb}.  Present
observations clearly indicate the potential for large spins on black
holes in binaries, possibly close to the Kerr limit $| {\bm{S}/{m}^2}
|=| {\bm{\chi}} |=1$~\cite{Gou:2011nq, Gou:2013dna, Fabian:2012kv,
  Nowak:2011wx, Reynolds:2011sba, Brenneman:2011wz}. Very few
measurements of the angles between the spins and the orbital angular
momentum exist from electromagnetic observations. In some cases, one
can measure this spin misalignment via \ac{GW} emission, as
misalignment leads to precession of the orbital plane,
which appears as phase and amplitude modulations in the observed
signal~\cite{Apostolatos:1994mx, Hannam:2013oca, Vitale:2014mka,
  O'Shaughnessy:2014dka}. However, for GW150914 it was not possible
to constrain the spin misalignment~\cite{TheLIGOScientific:2016wfe}.

We focus here on the effect of misalignment between the spin and the
orbital angular momenta from the perspective of \ac{GW} searches.  The
usual detection strategy for compact binaries is based on matched
filtering of the data against a bank of templates spanning as densely
as possible the full physical parameter
space~\cite{Allen:2005fk, Babak:2012zx}.
 Covering the full parameter space is quite
challenging. In fact, almost all searches of Initial
LIGO and Initial Virgo data for \acp{CBC} ignored
the effect of spins in search
templates~\cite{Abbott:2003pj, Abbott:2005kq, 
Abbott:2007xi, Abbott:2009tt, Abadie:2011kd, 
Colaboration:2011np, Aasi:2012rja,
  Briggs:2012ce}. The only search that did include spin effects
in its templates~\cite{Abbott:2007ai} was a search using
precessing-spin templates~\cite{Buonanno:2004yd}. This search was later
shown to perform, on average, no better than nonspinning searches~\cite{VanDenBroeck:2009gd}, as
the increased degrees of freedom in the signal space picked up extra
noise, which offset the gains achieved in signal-power
recovery. Several recent studies
performed with aligned-spin templates have demonstrated that it is possible
to pull in more signal than noise~\cite{Privitera:2013xza,
  Harry:2013tca, Canton:2014uja,Capano:2016dsf}, and aligned-spin
templates are currently used in searches with Advanced
LIGO~\cite{TheLIGOScientific:2016qqj}.  However, the effect of precession on the
GW signal is not fully captured by aligned-spin
templates~\cite{Ajith:2012mn,Brown:2012gs,Harry:2013tca},
and it remains to be seen whether precessing-spin effects can be
exploited to further improve the sensitivity of \ac{CBC} searches.
We note that while \emph{detection} searches currently consider
only aligned-spin templates, \emph{parameter inference} algorithms,
including the analysis of GW150914, do consider precessing waveforms~\cite{TheLIGOScientific:2016wfe,Veitch:2014wba}.
Studies already exist in the literature demonstrating that for systems
with strong precessional features it will be possible to break degeneracies
that exist between physical parameters in the emitted gravitational-wave
signal and therefore better measure the component spins, and potentially spin
orientations~\cite{Vitale:2014mka,O'Shaughnessy:2014dka}.
However, to be able to make such measurements we must
first be able to \emph{detect} highly precessing systems, and the detection
problem is the focus of this work.

In this work, we demonstrate a method for performing a detection search with
precessing-spin templates.  We restrict our
attention to \ac{BBH} and \ac{NSBH} systems, as neutron stars are generally
expected to have smaller spins compared to black holes~\cite{Lorimer:2008se,Chakrabarty:2008gz},
and precessing-spin binary neutron star signals already
match well with aligned-spin waveforms~\cite{Brown:2012qf,Ajith:2012mn}. 
More specifically, we construct template banks of
waveforms for precessing \ac{BBH} systems with component masses $m_i\geq
3M_\odot$, total masses $M_\mathrm{total}\in[6,100]M_\odot$ and mass
ratios $m_1/m_2 \leq 5$, as well as for precessing \ac{NSBH} systems
with component masses $3 M_\odot < m_1 < 15 M_\odot$, $1 M_\odot < m_2
< 3 M_\odot$.

Our method makes use of a new statistic, which maximizes the
matched-filter \ac{SNR} over the detector-sky location of the source in
addition to the phase and amplitude of the observed signal. We compare
our new method to searches using aligned-spin templates and quantify
the relative sensitivity between the two in a two-detector analysis of
simulated Advanced LIGO noise.  In doing so, we consider both the gain
in signal power due to having templates that more accurately model
precessing signals, as well as the increase in background from
filtering the data with more templates.  We find that, averaged across
the parameter space, these two factors largely cancel out, such that
the expected rate of observations at a fixed rate of false alarms with
the search presented here is roughly the same as that with an
aligned-spin-only search. However, in certain regions of parameter
space, namely at high mass ratios and large in-plane spins, we observe
an increase in observation rate that is greater than 50\%.

This paper is organized as follows. In Sec.~\ref{sec:waveforms}, we
define the \ac{BBH} and \ac{NSBH} parameter spaces considered in this
work, and justify the choice of waveform approximants used to model
these systems. In Sec.~\ref{sec:background}, we review the standard
formulation of the phase-, amplitude- and time-maximized \ac{SNR}
currently used in searches with nonprecessing templates, highlighting
the assumptions which are invalid for precessing signals. Having thus
laid out the mathematical formalism, we then derive in
Sec.~\ref{sec:prec_det_stats} a phase-, amplitude-, time- and
sky-location-maximized \ac{SNR} statistic applicable to precessing
templates, which we call the \ac{SMS}. We also present in
Sec.~\ref{sec:prec_det_stats} a comparison between the \ac{SMS} and
previous methods, indicating in particular the reasons for which we
find the sky-maxed \ac{SNR} approach to be more promising.  In
Sec.~\ref{sec:bank_cons}, we apply the \ac{SMS} to construct banks of
aligned-spin and precessing-spin template waveforms and demonstrate
their coverage of the precessing signal space. In
Sec.~\ref{sec:searches}, we apply these template banks in a real
pipeline analysis of simulated Gaussian noise in order to measure the
increase in background incurred by using precessing template banks
with the sky-maxed \ac{SNR}. From the measured increase in background
rate, we deduce the expected sensitivity improvement one could gain by
employing such a statistic in Gaussian noise. Finally, in
Sec.~\ref{sec:discussion}, we summarize the results, indicating the
work remaining to turn the method presented here into a truly viable
search method in real, non-Gaussian \ac{GW} data.

\section{Parameter Space and Waveform Models}
\label{sec:waveforms}

\begin{table}[tbp]
  \begin{tabular}{ccc}
  \begin{tabular}{cc}
    \multicolumn{2}{c}{\ac{BBH} parameter ranges} \\ \hline
    $m_1$, $m_2$ & $[3,97]~M_\odot$ \\
    $M_\mathrm{total}$ & $[6, 100]~M_\odot$ \\
    $m_1/m_2$ & $[1,5]$ \\
    $|\chi_1|$, $|\chi_2|$ & $[0, 0.99]$ \\
  \end{tabular}
  &
  &
  \begin{tabular}{cc}
    \multicolumn{2}{c}{NSBH parameter ranges} \\ \hline
    $m_1$ & $[3, 15]~M_\odot$ \\
    $m_2$ & $[1, 3]~M_\odot$ \\
    $|\chi_1|$ & $[0, 1.0]$ \\
    $|\chi_2|$ & $[0, 0.05]$ \\
  \end{tabular}
  \end{tabular}
    \caption{\label{tbl:mass_spin_params} Component mass and spin parameter
      ranges used to generate the waveforms used in this work for the BBH (left)
      and NSBH (right) parameter spaces. In both cases, the parameter space
      ranges are chosen identically for the template bank and the simulated signals
      used to test the coverage of the bank. For simulations, masses are drawn
      uniformly in component masses, spins are drawn uniformly in component spin
      magnitude, and all angular parameters are drawn isotropically. The
      parameter distribution for the stochastic placement of templates is
      detailed in the Appendix.}
\end{table}

In this paper, we consider two separate regions of the compact-binary
parameter space, shown in Table~\ref{tbl:mass_spin_params},
corresponding to \ac{NSBH} and \ac{BBH} sources. We use different
template waveform approximants in each region, based on the
considerations below.

Let us first discuss the NSBH parameter space.
For the purposes of detection with LIGO and Virgo, the signal
from compact binaries with $\mtot \lesssim 10-15 M_\odot$ are
well-modeled by the \ac{PN} approximation to the Einstein
field equations \cite{Blanchet:2013haa}, since only the inspiral portion of the signal is in band.
Truncation of the various physical ingredients (energy, flux, precession equations,
waveform amplitude) at different \ac{PN} orders, together with distinct methods for reexpanding
the balance equation when obtaining the frequency evolution equation, lead to a variety of
so-called \ac{PN} approximants, whose predictions for the signal can vary significantly~\cite{Buonanno:2009zt,Nitz:2013mxa}.
In this paper, our goal is not to compare these different models but rather to
understand the effect of adding precession to a search. While the details
will certainly depend on the approximant that one considers, it seems reasonable
to consider that the main effects will be captured by choosing one of them.

Specifically, when considering the NSBH parameter space, we use
the TaylorT2~\cite{Buonanno:2009zt} approximant with all orbital phase-evolution
terms up to 3.5PN order, all spin phase-evolution terms up to 2.5PN order, and using only the
dominant amplitude term. The waveform is generated from a frequency lower than that used
in the matched-filter integral and terminated at the frequency corresponding to the innermost 
stable circular orbit (ISCO) of a nonspinning black hole.
In certain cases, the evolution reaches a minimum energy configuration 
before the ISCO is reached, and the waveform terminates at that point if this happens.
For the purpose of bank placement, a faster way to evaluate waveform models is required. Thus,
for template bank construction only, we use the closed-form, single-spin, frequency-domain precessing model introduced
in Ref.~\cite{Lundgren:2013jla}. This waveform is derived from the TaylorT2
reexpansion of the balance equation and has been shown to agree well with TaylorT2 in most
cases~\cite{Lundgren:2013jla}.

For the larger masses considered in our BBH parameter space, the merger and ringdown portions of
the signal enter the detectors' most sensitive band, and including
these portions of the waveform becomes critical. Two approaches
have been developed over the past years for the construction of
approximate analytical models calibrated to numerical-relativity (NR) waveforms and
covering the entire coalescence of the binary. One approach is the effective-one-body (EOB) 
formalism \cite{Buonanno:1998gg,Buonanno:2000ef}, which combines a resummation of the available \ac{PN}
information and a description of the merger and ringdown phases with the calibration of a limited number of parameters 
(henceforth, EOBNR). The other approach is to construct phenomenological, frequency-domain models,
which directly interpolate between NR waveforms hybridized to \ac{PN} or EOB 
inspiral waveforms~\cite{Ajith:2009bn} (henceforth, IMRPhenom). Both approaches were successfully
applied to the simpler problem of modeling aligned-spin systems
\cite{Taracchini:2012ig,Taracchini:2013rva,Nagar:2015xqa,Damour:2014sva,Santamaria:2010yb,Husa:2015iqa,Khan:2015jqa}
and have recently been extended to the case of precessing systems \cite{Pan:2013rra,Hannam:2013oca}.

The template bank construction method that we use requires that the
computational cost of generating waveforms is small.  Precessing
time-domain EOBNR models are currently orders of magnitude too slow
for such a study; their use in these applications will require the
development of fast frequency-domain reduced-order surrogate models~\cite{Field:2013cfa},
like those already constructed for aligned-spin EOBNR
models~\cite{Purrer:2014fza,Purrer:2015tud}.  Therefore, we compute
template waveforms in our \ac{BBH} parameter space using a
phenomenological precessing
approximant~\cite{Hannam:2013oca,Schmidt:2014iyl}. This approximant is constructed from an underlying
aligned-spin model~\cite{Husa:2015iqa,Khan:2015jqa}, which models the
waveform in the coprecessing frame. Then, assuming a single in-plane
spin, \ac{PN} expressions are used to compute the precession
angles as a function of orbital frequency. The $\ell=2$ modes of
the waveform in an inertial frame are then obtained by appropriately
rotating the coprecessing aligned-spin waveform~\cite{Hannam:2013oca,
  Schmidt:2014iyl}.

This IMRPhenom model has known pathologies in the region of parameter
space where the projection of the total spin on the unit orbital angular
momentum is large and negative as these
configurations strongly violate one of the central assumptions in the model,
namely that the direction of the total angular momentum remains
approximately constant throughout the evolution~\cite{Smith:2016qas}.
A simple approximate and conservative
boundary for the pathological region is given by $\chi_1^\ell < -3/q$
where $q=m1/m2 >1$. As can be seen from this expression, this does not
affect systems with $q<3$ but covers an increasingly larger portion of
the spin parameter space as $q$ increases. 
For this reason, we decide to restrict our parameter space to $q\leq5$,
for which the problematic portion of parameter space (determined using
the proper condition involving also the projection of the spin of the
secondary body) is only a small fraction ($\simeq 0.3\%$) of the full
space. By this we mean that assuming the distributions described
in the caption of Table.~\ref{tbl:mass_spin_params}, only $0.3\%$ of the simulations lie in
this region. In the rest of this analysis, we excise this region from our
parameter space both when assessing sensitivity improvement via sets of
simulations and when constructing our template banks. This is
also the reason why we do not use this model in the NSBH region of parameter
space despite the obvious advantage of including the merger-ringdown portion
and therefore removing the uncertainties related to the termination of the
waveform.

All the waveforms that we use in this paper are publicly available
in the \texttt{lalsimulation} repository~\cite{LAL}\footnote{The
  internal \texttt{lalsimulation} names for the waveforms described
  above are ``SpinTaylorT2'' and ``SpinTaylorF2'' for the two PN
  NSBH models,  ``SEOBNRv2'' for the aligned-spin EOBNR approximant,
  ``SEOBNRv3'' for the precessing EOBNR approximant,
  ``IMRPhenomPv2'' for the precessing phenomenological approximant and
  ``IMRPhenomD'' for its aligned-spin counterpart.}.

\section{Sky-maxed SNR: NonPrecessing Limit }
\label{sec:background}

We now lay out the  mathematical formalism we later use for
deriving our \ac{SMS} by first reviewing the phase-maximized matched
filter \ac{SNR} for non-precessing templates. This method has been used in
nearly all initial LIGO and Virgo \ac{CBC} searches to
date~\cite{Abbott:2003pj, Abbott:2005kq, Abbott:2007xi,
Abbott:2009tt, Abadie:2011kd,
  Colaboration:2011np, Aasi:2012rja, Allen:2005fk}
and is well described in existing literature~\cite{Thorne300,Sathyaprakash:1991mt,Allen:2005fk,Babak:2012zx}.
  
Consider the data output $s(t)$ of a
\ac{GW} detector, which consists of noise $n(t)$ and
possibly a \ac{GW} signal of known form $h(t)$. We wish to
decide between the signal hypothesis and the noise hypothesis,
\[ s(t) = \begin{cases} 
        n(t) & \textrm{noise hypothesis} \\
       n(t) +  h(t) & \textrm{signal hypothesis}
    \end{cases},
\]
given the observed data and predicted form of the signal. We assume
that the noise is both stationary and Gaussian. Under these
assumptions, the statistical properties of the noise are fully
described by a single function, the one-sided ($f>0$) noise \ac{PSD}
$S_n(f)$, defined by
\begin{equation}
 \frac{1}{2} \delta(f-f') S_n(f) = \mathbb{E}[ \tilde{n}(f) \tilde{n}^{*}(f')], 
\end{equation}
where $\mathbb{E}[\cdot]$ denotes the expectation value over
independent noise realizations. The \ac{PSD} naturally induces a
complex inner product $\langle | \rangle$ on the signal space,
\begin{equation}
 \langle a|b\rangle = 4 \int^\infty_0 \frac{\tilde{a}(f)\tilde{b}^*(f)}{S_n(f)} df.
\end{equation}
Using the Gaussian assumption, we can then express the probabilities
$P(s|n)$ and $P(s|h)$ of the observed data given the signal and noise
hypotheses, respectively, in terms of this inner product as
\begin{eqnarray}
 P(s|n) &\propto& e^{- \Re[ \left\langle s | s\right\rangle]/2 } \\
 P(s|h) &\propto& e^{- \Re[ \left\langle s-h | s-h\right\rangle]/2 },
\end{eqnarray}
where $\Re$ denotes the real part. It follows that the likelihood
ratio $\Lambda \equiv P(s|h)/P(s|n)$ between the signal and noise
hypotheses is given by
\begin{equation}
  \log \Lambda \equiv \lambda = \Re\left[ \left\langle s | h \right\rangle \right]
      - \frac{1}{2} \Re \left[ \left\langle h | h \right\rangle \right].
  \label{eqn:matched_filter_snr_raw}
\end{equation}
By the Neymann-Pearson lemma, a search which thresholds on the
matched-filter statistic given in Eq.~\eqref{eqn:matched_filter_snr_raw}
maximizes the probability of accepting the signal hypothesis whenever
the signal hypothesis is true for any false-alarm
probability. Equation~\eqref{eqn:matched_filter_snr_raw} therefore gives
the general prescription for searching for a \ac{GW} signal of known
form in stationary, Gaussian noise. One also defines the matched-filter SNR, $\rho$, by maximizing $\lambda$ over
an overall amplitude
\begin{equation}
 \label{eqn:matched_filter_snr_maxamp}
 \rho^2 / 2 = \max_{\textrm{amp}} (\lambda) = \frac{\left( \Re \left[ \left\langle s | h \right\rangle \right] \right)^2}
    {\left\langle h | h \right\rangle} = \left(\Re [ \langle s | \hat{h} \rangle ]\right)^2,
\end{equation}
where $\hat{x} \equiv x / \langle x | x \rangle^{1/2}$ is used to denote normalized waveforms.

In practice, the exact form of the signal is not known, and we must
maximize Eq.~\eqref{eqn:matched_filter_snr_raw} over the {\it a
  priori} unknown template parameters which determine the signal.  A
generic compact-binary coalescence \ac{GW} signal is described by at
least fifteen parameters\footnote{We restrict
  attention to compact binaries on circular orbits, removing
  parameters related to eccentricity. We also ignore any effects
  related to the internal structure of neutron stars.}: the component
masses, $m_1$ and $m_2$; the component dimensionless spin vectors
$\bm{\chi}_1$, $\bm{\chi}_2$, the sky location of the signal with
respect to the frame of the observer ($\theta$, $\phi$); the distance
$D$ to the source; the coalescence time $t_c$ of the signal; the
inclination of the binary with respect to the line-of-sight to the
system $\iota$; a polarization angle $\psi$; and an orbital phase at
coalescence $\phi_c$.  For some parameters, this maximization step can
be performed analytically, or in a computationally efficient way using
\ac{FFT} algorithms, whereas for the remaining ones, one has to resort
to discretizing the parameter space and repeating the matched-filter
operation a large number of times. In this and the following sections,
we are concerned with the possible analytic maximizations of
Eq.~\eqref{eqn:matched_filter_snr_raw}. In Sec.~\ref{sec:bank_cons},
we describe how we create banks of waveforms to optimize over the
remaining parameters.

The observed signal $h(t)$ at the detector is the
sum of the two \ac{GW} polarizations, $h_+$ and $h_\times$,
multiplied by the response function of the detector to each
polarization, $F_+$ and $F_\times$~\cite{Finn:1992xs}, which
encapsulate the full dependence of the signal on $(\theta, \phi,
\psi)$
\begin{equation}
  h(t) = F_+(\theta, \phi, \psi)\, h_+(t) + F_{\times}(\theta, \phi, \psi)\,h_{\times}(t).
   \label{eqn:fplusfcross}
\end{equation}
It is this combination of $h_+$ and $h_\times$ given in
Eq.~\eqref{eqn:fplusfcross} that is used as the filter of
Eq.~\eqref{eqn:matched_filter_snr_raw}. 

In making the connection between the two waveform polarizations and the template waveform $h(t)$, aligned-spin searches rely
on the simplifying assumption that only the dominant $(\ell,|m|)=(2,2)$ modes
of the waveform\footnote{Defining these modes in a natural radiation
  frame where the $z$-axis, with respect to which the multipolar
  decomposition is performed, coincides with the direction of the 
  angular orbital momentum.}
are taken into account. This allows us to write the dependence of the two \ac{GW} polarizations on the physical parameters in the form~\cite{Babak:2012zx}
\begin{eqnarray}
    h_{+} &=& \frac{1 + \cos^2\iota}{2D} A(t-t_c;\xi) \cos[2(\Phi(t-t_c;\xi)+\phi_c)], \nonumber \\
    h_\times &=& \frac{\cos \iota}{D} A(t-t_c;\xi) \sin[2(\Phi(t-t_c;\xi)+\phi_c)],
    \label{eqn:hpctalignedspins}
\end{eqnarray}
where $A(t;\xi)$ and $\Phi(t;\xi)$ are functions of time and the
parameters $\xi=(m_1,m_2,\chi_1,\chi_2)$, where $\chi_1$ and
$\chi_2$ denote the constant projections of the spins in the direction
of the orbital angular momentum. Inserting
Eq.~\eqref{eqn:hpctalignedspins} into Eq.~\eqref{eqn:fplusfcross}, we
find that the full strain seen by a detector can now be written as
\begin{equation} \label{eqn:hoft_alignedspin}
  h(t) = \frac{A(t-t_c;\xi)}{D_\mathrm{eff}} \cos[2(\Phi(t-t_c;\xi)+\phi_0)],
\end{equation}
where
\begin{equation}
D_\mathrm{eff} = D \left[ F_+^2\left(\frac{1 + \cos^2 \iota}{2}\right)^2 + F_\times^2 \cos^2 \iota \right]^{-1/2}
\end{equation}
is the so-called effective distance and $\phi_0$, defined as
\begin{equation}
e^{2 i \phi_0} = e^{2 i \phi_c} \frac{F_+ (1 + \cos^2 \iota)/2 - i F_\times \cos \iota }{\left[ F_+^2\left(\frac{1 + \cos^2 \iota}{2}\right)^2 + F_\times^2 \cos^2 \iota \right]^{1/2}},
\end{equation}
is the phase of the observed waveform at coalescence.
Thus, in the aligned-spin case, the waveform $h(t)$ depends on the
parameters $\phi_c$, $D$, $\theta$, $\phi$, $\psi$ and $\iota$ only
through the combinations $D_\mathrm{eff}$ and $\phi_0$. Moreover, this
dependence amounts only to an overall phase and an overall amplitude.

For search applications, we ultimately need the dependence of the Fourier
transform of $h(t)$ on the physical parameters. Assuming that the time scale
over which the amplitude $A$ changes is much smaller than the orbital
time scale we can apply the stationary phase approximation~\cite{Allen:2005fk}.
This allows us to conveniently factorize the dependence on
$\phi_0$ as
\begin{equation} \label{eqn:aligned_spin_fd_filter}
  \tilde{h}=\frac{1}{D_\mathrm{eff}} e^{2i\phi_0} \tilde{h}_0(f;t_c, \xi),
\end{equation}
where we have defined
\begin{equation} 
  h_0(t-t_c,\xi)=A (t-t_c;\xi) \cos[2(\Phi(t-t_c;\xi)],
\end{equation}
which depends only on $\xi=(m_1,m_2,\chi_1,\chi_2)$ and $t_c$.

We can now maximize Eq.~\eqref{eqn:matched_filter_snr_raw} over the parameters
$\phi_c$, $D$, $\theta$, $\phi$, $\psi$, and $\iota$ by maximizing
over the combinations of parameters $D_\mathrm{eff}$ and $\phi_0$.
Inserting Eq.~\eqref{eqn:aligned_spin_fd_filter} into
Eq.~\eqref{eqn:matched_filter_snr_raw} and maximizing with respect to
$D_\mathrm{eff}$ and $\phi_0$, we thus obtain
\begin{equation} \label{eqn:max_likelihood_kphi}
 \max_{\iota,D,\theta,\phi,\psi,\phi_c}\left( \lambda \right) =
 \frac{1}{2}\max_{\phi_0}\left(\rho^2\right) = 
 \frac{1}{2} \lvert \langle s| \hat{h}_0\rangle\rvert^2.
\end{equation}

The coalescence time $t_c$ parametrizes time translations of $h_0$.
Therefore in the Fourier domain we can write
$\tilde{h}_0(f; t_c,\xi)=\tilde{h}_0(f; \xi)e^{-2i\pi f t_c}$
and therefore
\be
\label{eqn:tcmaximizationalignedspins}
\langle s|h_0\rangle= \int^\infty_0 \frac{\tilde{s}^*(f)
  \tilde{h}_0(f;\xi)}{S_n(f)}e^{-2 \pi i f t_c} df.
\ee
Evaluating $\langle s|h_0\rangle$ over a range of $t_c$
can be efficiently performed
numerically by using widely available \ac{FFT} routines.

To summarize, it is possible in the aligned-spin case to quickly
maximize over all parameters describing the system except for
$\xi=(m_1, m_2, \chi_1, \chi_2)$, provided subdominant modes can be
neglected and that the stationary-phase approximation holds. The
remaining parameters are searched over by repeating the matched-filter
operation Eq.~\eqref{eqn:max_likelihood_kphi} over a discrete bank of
templates, with time maximization handled by an efficient FFT
implementation of Eq.~\eqref{eqn:tcmaximizationalignedspins}. We
discuss the construction of the discrete template banks in
Sec.~\ref{sec:bank_cons}.

\section{Sky-maxed SNR: Precessing limit}
\label{sec:prec_det_stats}

Consider now the case where we wish to conduct a search using waveforms with generically oriented spins.
A first obvious difference with the case described above is that we now have to deal
with the six dimensionless spin components: $\bm{\chi}_1$ and $\bm{\chi}_2$\footnote{We do
not need to specify here the frame used to define the spin components;
note however that as the spins evolve with time in the precessing
case, we define these values to be the spins at some reference time.}.
A more important difference is that in the precessing case the
orientation of the source with respect to the detector varies as the
orbit precesses.  As a result, the two polarizations $h_+$ and
$h_\times$ cannot be written in the simple form of
Eq.~\eqref{eqn:hpctalignedspins} where both are identical up to an
amplitude rescaling that only depends on $\iota$ and a constant phase
shift. Therefore, we return to Eq.~\eqref{eqn:fplusfcross} and derive
a new statistic free of this assumption on the waveform. This
statistic maximizes $\lambda$, not only over an amplitude and a phase,
but also over the sky-location-dependent antenna factors. We then
explore the statistical properties of this maximized form of $\lambda$
and compare to previously proposed approaches, emphasizing the
differences which make the current approach more promising.

\subsection{Sky-maximized signal-to-noise ratio}
\label{ssec:prec_stat}

We start by expressing the dependence in Eq.~\eqref{eqn:fplusfcross} on the detector
related angles $(\theta,\, \phi, \psi)$ and the distance $D$  in terms of an
overall amplitude and a phase between $h_+$ and $h_{\times}$,
\bea
\label{eqn:hTDprecessing}
h=K(\theta,\phi,\psi,D)\Big[h_+(t;t_c,\xi,\iota,\phi_c) \cos \kappa(\theta,\phi,\psi&)\nonumber\\
+ h_\times(t;t_c,\xi,\iota,\phi_c) \sin \kappa(\theta,\phi,\psi)&\Big],
\eea
where we have defined $\xi=(m_1,m_2,\bm{\chi}_1,\bm{\chi}_2)$ and
\bea
& \displaystyle e^{i \kappa} = \frac{F_+ + i F_\times}{\sqrt{F_+^2+F_\times^2}},\\
& \displaystyle K = \frac{1}{D}\sqrt{F_+^2+F_\times^2}.
\eea
In addition, as in Eq.~\eqref{eqn:aligned_spin_fd_filter}, we can factorize the
dependence of $\phi_c$ in the Fourier domain as
\be
\tilde{h}=K e^{2i\phi_c} [\tilde{h}_+(\phi_c=0) \cos \kappa +\tilde{h}_\times(\phi_c=0) \sin \kappa ].
\label{eq:fgenericFourier}
\ee
As with the aligned-spin waveforms, this is not an exact symmetry. However,
it is a particularly good approximation if one only considers a waveform
containing the $(\ell,|m|)=(2,2)$ modes in the corotating
frame~\cite{Buonanno:2004yd,Hannam:2013oca}.

We could now perform the maximization over the amplitude $K$ and the
phase $\phi_c$ just as in Eq.~\eqref{eqn:max_likelihood_kphi} but with
$h_0$ replaced by $\tilde{h}_+ \cos \kappa +\tilde{h}_\times \sin
\kappa$. This would leave us with the two additional parameters
$\iota$ and $\kappa$ (plus the four new spin components) to be covered
using a discrete bank.  While it is possible to construct template
banks in this manner, it would be desirable to further
reduce the dimension of parameter space.  Furthermore, different
\ac{GW} observatories, with different orientations and locations, will
not observe the \emph{same} combination of sky angles $\kappa$.  We
therefore consider a scheme where we maximize not only over the
overall amplitude $K$ and phase $\phi_c$, but also over the angle
$\kappa$. Such a scheme removes all detector-dependent quantities from
the parameters used when constructing the template bank, allowing the
use of a simpler template coincidence method for a multidetector
analysis.

Maximizing the log-likelihood defined in Eq.~\eqref{eqn:matched_filter_snr_raw} over $K$ and $\phi_c$, straightforwardly leads to 
\begin{equation} \label{eqn:likelihood_phase_amp_maxed_001}
 \max_{K,\phi_c} (\lambda) =
  \frac{1}{2} \frac{u^2 |\hat{\rho}_+ |^2
   + 2 u \hat{\gamma}
   + |\hat{\rho}_{\times} |^2}{u^2 + 2 u I_{+\times} + 1},
\end{equation}
where we have defined
\begin{eqnarray}
 \hat{\rho}_{+,\times} &=& \langle s | \hat{h}_{+,\times} \rangle \\
 \hat{\gamma} &=& \Re\left[ \hat{\rho}_+ \hat{\rho}^*_{\times} \right] \\
 \langle \hat{h}_+ | \hat{h}_{\times} \rangle &=& I_{+ \times} + i J_{+\times} \label{eqn:defI+x},
\end{eqnarray}
with $I_{+ \times}, J_{+ \times} \in \mathbb{R}$
and we factorize the $\kappa$ dependence in terms of
\begin{equation}
  u \equiv \frac{1}{\tan \kappa} \sqrt{ \frac{\langle h_+ | h_+ \rangle }{\langle h_\times | h_\times \rangle} }.
\end{equation}

\begin{widetext}
Taking the derivative of Eq.~\eqref{eqn:likelihood_phase_amp_maxed_001}
with respect to $u$ and solving for the roots leads to a quadratic
equation in $u$,
\begin{equation}
 (I_{+\times} |\hat{\rho}_+|^2 - \hat{\gamma}) u^2
 + (|\hat{\rho}_+|^2 - |\hat{\rho}_{\times}|^2) u
 + (\hat{\gamma} - I_{+\times} |\hat{\rho}_\times|^2) = 0.
\end{equation}
Substituting the roots of this equation back into
Eq.~\eqref{eqn:likelihood_phase_amp_maxed_001}, we obtain two extremal
values for $\lambda$,
\begin{equation} \label{eqn:likelihood_phase_amp_u_maxed_002}
  \lambda = \frac{1}{4}\left( \frac{|\hat{\rho}_{+}|^2 - 2 \hat{\gamma} I_{+\times} + |\hat{\rho}_\times|^2
      \pm \sqrt{(|\hat{\rho}_+|^2 - |\hat{\rho}_\times|^2)^2 + 4 (I_{+\times} |\hat{\rho}_+|^2 - \hat{\gamma})
              (I_{+\times} |\hat{\rho}_\times|^2 - \hat{\gamma}) }}
  {1 - I_{+\times}^2} \right).
\end{equation}
To take the maximal value of $\lambda$, we notice that the denominator
of Eq.~\eqref{eqn:likelihood_phase_amp_u_maxed_002} is always positive, so the log-likelihood will always take a maximum value
when the square-root term is positive. Therefore,
\begin{equation} \label{eqn:likelihood_phase_amp_u_maxed_003}
   \max_{D,\phi_c,\theta, \phi, \psi}(\lambda) = \max_{K,\phi_c,u}(\lambda) = \frac{1}{4}\left( \frac{
      |\hat{\rho}_{+}|^2 - 2 \hat{\gamma} I_{+\times} + |\hat{\rho}_\times|^2
      + \sqrt{(|\hat{\rho}_+|^2 - |\hat{\rho}_\times|^2)^2 + 4 (I_{+\times} |\hat{\rho}_+|^2 - \hat{\gamma})
              (I_{+\times} |\hat{\rho}_\times|^2 - \hat{\gamma}) }}
  {1 - I_{+\times}^2} \right),
\end{equation}
\end{widetext}
is the log-likelihood maximized over an overall phase, an overall
amplitude and the sky location of the binary. 

We notice that Eq.~\eqref{eqn:likelihood_phase_amp_u_maxed_003} is ill defined
in the case that $I_{+\times} = \pm1$. However, for compact-binary waveforms
it is not possible for the $\hat{h}_+$ and $\hat{h}_{\cross}$ components to be identical
and so this case can never occur. Additionally, the terms within the
square root of Eq.~\eqref{eqn:likelihood_phase_amp_u_maxed_003} will
always take positive values and therefore the equation will always
produce real, positive, values of $\lambda$.
In analogy with the
nonprecessing case discussed in the previous section, we define
\begin{equation} \label{eq:sms}
  \max_{K,\phi_c,u} (\lambda) = \frac{1}{2}\max_{\phi_c,u}\left(\rho^2\right)
  = \frac{1}{2}\rho^2_{\mathrm{SM}} ,
\end{equation}
and we refer to the quantity $\rho_{\mathrm{SM}}$ as the \ac{SMS}.

Additionally, Eq.~\eqref{eqn:likelihood_phase_amp_u_maxed_003} can be
maximized over $t_c$ in a similar way as in the aligned-spin case:
both $\hat{\rho}_{+,\times}$ can be efficiently computed for a
discrete set of values of $t_c$ using \ac{FFT} algorithms and one can
just pick the largest resulting value of $\lambda$. Note, however, that
unlike in the aligned-spin case, maximizing the likelihood over the unknown 
coalescence time in the precessing case requires the computation of
\emph{two} inverse \acp{FFT}, which contributes to increasing the
computational cost.

As a sanity check, we show how the \ac{SMS} behaves in
the aligned-spin limit. In this case, the simple
relation between the polarizations Eq.~\eqref{eqn:hpctalignedspins} implies that in the frequency
domain
\begin{equation}
 \hat{h}_+ = \pm i \hat{h}_{\times},
\end{equation}
and therefore
\begin{eqnarray}
 \hat{\rho}_+ &=& \pm i \hat{\rho}_{\times}, \\
 I_{+ \times} &\equiv& \Re \left[ \langle \hat{h}_+ | \hat{h}_{\times} \rangle \right] = 0.
\end{eqnarray}
Inserting these conditions into Eq.~\eqref{eqn:likelihood_phase_amp_u_maxed_003} results
in the equation collapsing to the form of Eq.~\eqref{eqn:max_likelihood_kphi},
\begin{equation}
\label{eqn:precessingstatisticcollapsedalignedspins}
\max_{D,\phi_c,\theta, \phi, \psi}(\lambda) = 
     \frac{1}{2} |\langle s | \hat{h}_+ \rangle |^2.
\end{equation}
This is of course expected: in this case, the sky location enters only as a constant phase and
amplitude shift and therefore by maximizing over these overall degrees of freedom,
one has already maximized over $\kappa$.

\subsection{Statistical distribution of the sky-maximized SNR in Gaussian noise}
\label{ssec:stat_distr}

In Eqs.~\eqref{eqn:likelihood_phase_amp_u_maxed_003} and~\eqref{eq:sms}, we have defined a new statistic to be used in
searches with precessing templates. This statistic has different
statistical properties, in general, than the standard aligned-spin
statistic defined in Eq.~\eqref{eqn:matched_filter_snr_maxamp}. Before
applying the \ac{SMS}, we wish to better understand these
differences. Here, we investigate the distribution of the (squared)
\ac{SMS} $\rho^2_{\mathrm{SM}}$ in Gaussian noise.

For a given template with a known value of $I_{+\times}$,
the statistic defined in Eq.~\eqref{eqn:likelihood_phase_amp_u_maxed_003} is
a combination of two complex variables, $\hat{\rho}_+$ and $\hat{\rho}_{\times}$. We use the
following notation for their real and imaginary parts,
\be
\hat{\rho}_{+,\times}=\epsilon_{+,\times}^R+i\epsilon_{+,\times}^I,
\ee
which are, for any point in time and in Gaussian noise, real Gaussian random variables with
unit variance and zero mean. Thus,
\begin{eqnarray}
 \mathbb{E}[\epsilon_+^R] = \mathbb{E}[\epsilon_+^I] &=&
 \mathrm{E}[\epsilon_{\times}^R] = \mathbb{E}[\epsilon_{\times}^I] = 0, \\
  \mathrm{Var}[\epsilon_+^R] = \mathrm{Var}[\epsilon_+^I] &=&
 \mathrm{Var}[\epsilon_{\times}^R] = \mathrm{Var}[\epsilon_{\times}^I] = 1.
\end{eqnarray}
However, while by definition the imaginary and real components of $\hat{\rho}_+$
(and separately of $\hat{\rho}_{\times}$) are statistically independent,
\begin{eqnarray}
 \mathrm{E}[\epsilon_+^R\epsilon_+^I] =
  \mathrm{E}[\epsilon_\times^R\epsilon_\times^I] = 0,
\end{eqnarray}
the correlation between $\hat{\rho}_+$ and $\hat{\rho}_{\times}$ will depend on the
template and \ac{PSD} being used.

In the aligned-spin case, we see directly from Eq.~\eqref{eqn:precessingstatisticcollapsedalignedspins}
that $2 \lambda$ is the sum of two independent Gaussian variables squared and therefore has a $\chi^2$ distribution
with 2 degrees of freedom. In the generic case, the statistical distribution will depend on both the real
and imaginary parts of $\langle \hat{h}_+ | \hat{h}_{\times} \rangle$: $I_{+\cross}$ and $J_{+,\cross}$.
In order to explore this, it is more convenient to reexpress
Eq.~\eqref{eqn:likelihood_phase_amp_u_maxed_003} in terms of four \emph{independent} standard normal
variables. We therefore introduce a normalized linear
combination,
\begin{equation}
\hat{h}_{\perp} = \frac{\hat{h}_{\times} - \langle \hat{h}_+ | \hat{h}_{\times } \rangle \hat{h}_+}
                       {\sqrt{1 - | \langle \hat{h}_+ | \hat{h}_{\times} \rangle |^2}},
\end{equation}
such that 
$\langle \hat{h}_+ | \hat{h}_{\perp} \rangle = 0$. From this we obtain a new complex variable
$\hat{\rho}_\perp$ whose real and imaginary parts $\epsilon_\perp^{R,I}$ are standard normal
variables statistically independent of $\epsilon_+^{R,I}$. Reexpressing $\hat{\rho}_{\times}$
as a linear combination of $\hat{\rho}_+$ and $\hat{\rho}_{\perp}$,
\begin{equation}
 \hat{\rho}_{\times} =  \langle \hat{h}_+ | \hat{h}_{\times } \rangle \hat{\rho}_+ +
         \sqrt{1 - | \langle \hat{h}_+ | \hat{h}_{\times} \rangle |^2} \hat{\rho}_{\perp},
\end{equation}
and plugging this into Eq.~\eqref{eqn:likelihood_phase_amp_u_maxed_003}, we rewrite the
statistic in terms of four statistically independent, zero-mean, unit-variance
vectors and the real and imaginary components of $\langle \hat{h}_+ | \hat{h}_{\times} \rangle$,
\begin{widetext}
\bea
\label{eqn:explicitlambdawithnormalvars}
\lambda &= &\Bigg[ \Bigg. 
H_{-}^2 |\hat{\rho}_{+}|^2
+H_{+}^2 |\hat{\rho}_{\perp}|^2
-2 J_{+\times} H_{+} (\epsilon_+^I\epsilon_\perp^R-\epsilon_+^R\epsilon_\perp^I)
+\Bigg(
H_{+}^4\Big[|\hat{\rho}_{+}|^4+|\hat{\rho}_{\perp}|^4\Big] \nonumber\\
&&+ 2 J_{+\times} H_{+}^3 (\epsilon_+^I\epsilon_\perp^R-\epsilon_+^R\epsilon_\perp^I)\big(|\hat{\rho}_{+}|^2- |\hat{\rho}_{\perp}|^2\big)
+H_{+}^2\bigg\{
8J_{+\times}^2(\epsilon_+^I\epsilon_\perp^R-\epsilon_+^R\epsilon_\perp^I)^2 \nonumber\\
&&+2 H_{-}^2\Big((\epsilon_+^R\epsilon_\perp^R)^2-(\epsilon_+^I\epsilon_\perp^R)^2+
4\epsilon_+^R\epsilon_+^I\epsilon_\perp^R\epsilon_\perp^I-(\epsilon_+^R\epsilon_\perp^I)^2+
(\epsilon_+^I\epsilon_\times^I)^2\Big)
\bigg\}\Bigg)^{1/2}\Bigg]/\left(4(1-I_{+\times}^2)\right).
\eea
\end{widetext}
Here we have defined for convenience
\bea
H_{-}=&\sqrt{1-(I_{+\times}^2-J_{+\times}^2)},\\
H_{+}=&\sqrt{1-(I_{+\times}^2+J_{+\times}^2)},
\eea
and we remind the reader for completeness that
\be
|\hat{\rho}_{+,\perp}|^2=(\epsilon_{+,\perp}^R)^2+(\epsilon_{+,\perp}^I)^2.
\ee

This fully explicit form, although not very elegant, allows us to easily identify some particular
cases and symmetries. First, as already discussed, in the aligned-spin case for which
$J_{+\times}=\pm1$, $I_{+\times}=H_{+}=0$ and $H_{-}=\sqrt{2}$, the distribution trivially collapses to a
$\chi^2$ distribution with 2 degrees of freedom $2\lambda=(\epsilon_{+}^R)^2+(\epsilon_{+}^I)^2$.
Another interesting case arises when $J_{+\times}=0$ since in this case $H_{-}^2=H_{+}^2=1-I_{+\times}^2$
and the dependence on $I_{+\times}$ completely cancels out. Furthermore, Eq.~\eqref{eqn:explicitlambdawithnormalvars}
allows us to show that the distribution of $\lambda$ does not depend on the sign of $I_{+\times}$ and
$J_{+\times}$. Indeed, Eq.~\eqref{eqn:explicitlambdawithnormalvars} is left invariant by the
transformation $I_{+\times}\rightarrow - I_{+\times}$ or by the transformation
$(J_{+\times},\epsilon_{\perp}^R,\epsilon_{\perp}^I)\rightarrow (-J_{+\times},-\epsilon_{\perp}^R,-\epsilon_{\perp}^I)$. 

\begin{figure}[tbp]
  \includegraphics[width=\columnwidth]{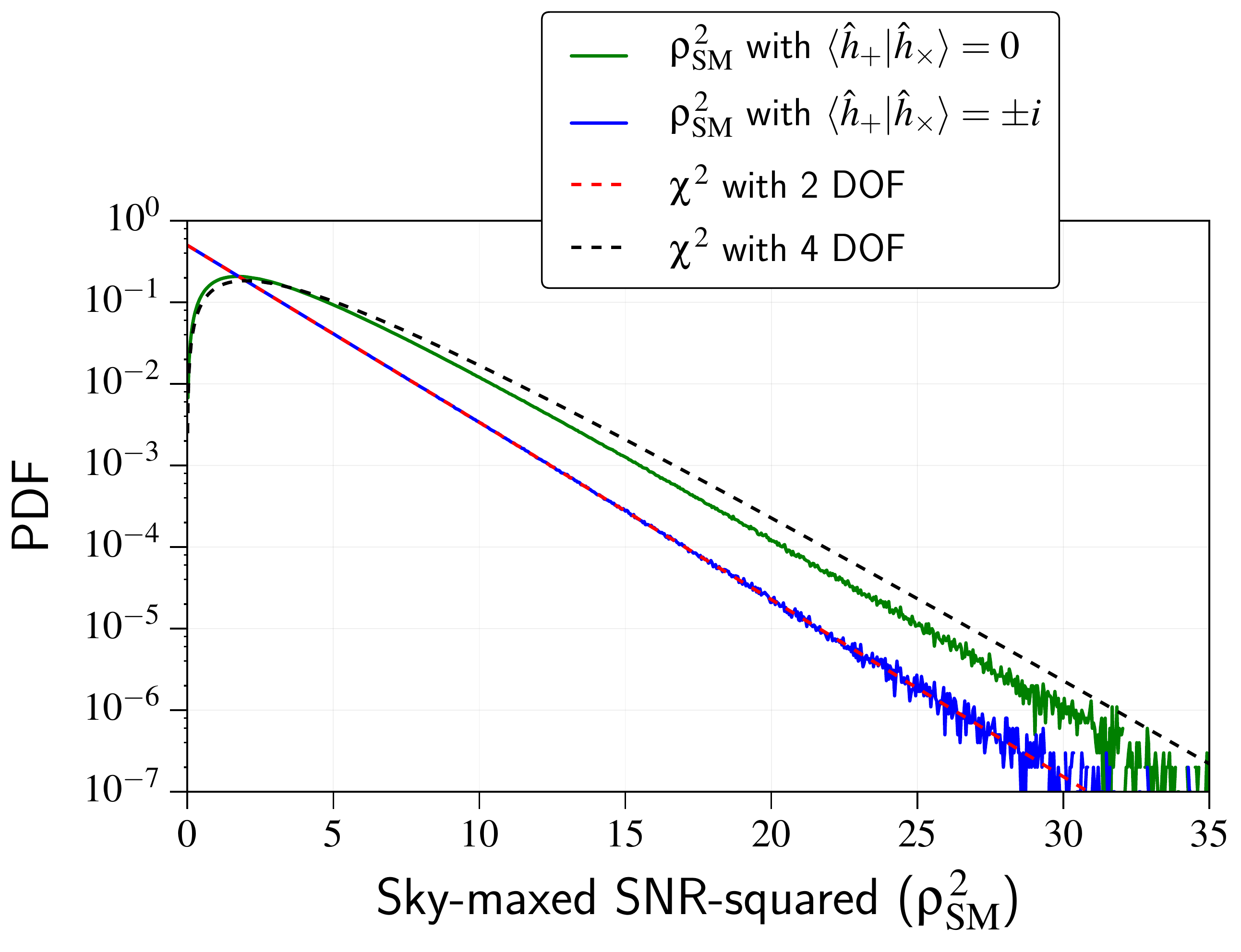}
  \caption{\label{fig:statdetfigure} The two limiting cases,
    ($\langle \hat{h}_+ | \hat{h}_{\cross}\rangle = 0$) and ($\langle \hat{h}_+ |
    \hat{h}_{\cross}\rangle = \pm i$), of the probability density function
    of the precessing \ac{SMS} in Gaussian noise.  Also
    plotted are $\chi^2$ distributions with 2 and 4 degrees of freedom
    for direct comparison. As expected, when $\langle \hat{h}_+ |
    \hat{h}_{\cross}\rangle = 0$ the statistic follows a $\chi^2$
    distribution with 2 degrees of freedom.}
\end{figure}

As we have demonstrated, $\lambda$ is formed from a combination of four orthogonal time
series which, in Gaussian noise will each be independent, and follow a normal distribution with
zero mean and unit variance. We identify an upper bound on the \ac{PDF}
of the \ac{SMS} at large \acp{SNR}
by considering the case where one is free to capture the power
in all four of these vectors, i.e. when the four components are added in quadrature.
In this case 2$\lambda$ would
follow a $\chi^2$ distribution with four degrees of freedom.
However, $\lambda$ as defined in Eq.~\eqref{eqn:explicitlambdawithnormalvars}
does not have the freedom to capture the power
in all four of these vectors, it is constrained to the physical subspace. We find
numerically that the \ac{PDF} of the \ac{SMS} takes the largest values
at high \acp{SNR} when $\langle \hat{h}_+ | \hat{h}_{\times} \rangle = 0$.
The \ac{PDF} takes the smallest values at high \acp{SNR}
in the case where $\langle \hat{h}_+ | \hat{h}_{\times}\rangle = \pm i$,
as for nonspinning or aligned-spin restricted waveforms.
This can be seen in Fig.~\ref{fig:statdetfigure}
where we compare the distribution of $2\lambda=\rho^2_{\mathrm{SM}}$ for both
the lower and upper bound configurations with $\chi^2$ distributions with 2,4 and 6 degrees
of freedom. We note that the lower bound at high SNRs for the ``PTF'' approach,
which we discuss in the next subsection, follows a $\chi^2$ distribution with 6 degrees of freedom.

\begin{figure*}[tbh]
  \includegraphics[width=2\columnwidth]{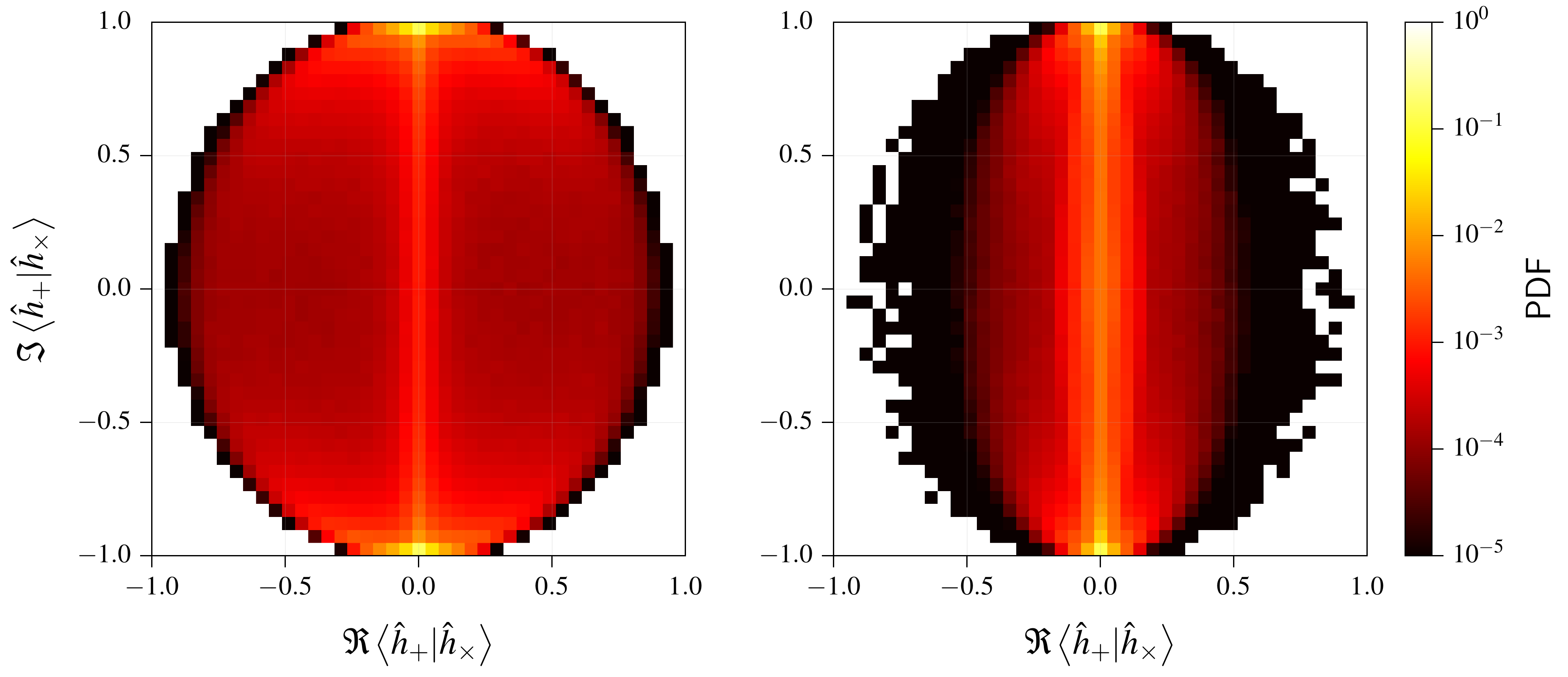}
  \caption{The distribution of $\langle \hat{h}_+ | \hat{h}_{\cross}\rangle$ for
    (left) a set of BBH signals modelled with the PhenomP approximant
    and (right) a set of NSBH signals modeled with the TaylorT2
    approximant, each set drawn with the distribution described in
    Sec.~\ref{ssec:stat_distr}.  Most waveforms cluster around the
    $\langle \hat{h}_+ | \hat{h}_{\cross}\rangle=\pm i$ value achieved exactly by
    aligned-spin systems.}
 \label{fig:tempcorrfigurev2}
\end{figure*}

Since the distribution of our \ac{SMS} depends on the value of
$\langle \hat{h}_+ | \hat{h}_{\times} \rangle$ for the template that one is considering,
we find it informative to visualize the distribution of $\langle \hat{h}_+ | \hat{h}_{\times} \rangle$
corresponding to a set of precessing waveforms randomly drawn from our parameter spaces
of interest, as shown in Fig.~\ref{fig:tempcorrfigurev2}. The particular distribution that the
simulations in the panels of this figure are drawn from are listed in Table~\ref{tbl:mass_spin_params}.
We see that even though the set of aligned-spin waveforms is
of measure zero, the distribution is highly peaked around the aligned-spin value $\pm i$.

\subsection{Comparison to previously proposed approaches}
\label{ssec:prec_det_stat_history}

The first attempt to derive an efficient search for
precessing waveforms in \ac{GW} data analysis was presented
in Ref.~\cite{Apostolatos:1995pj}. This approach involved adding an additional
free parameter at the same order as the 1PN orbital term,
to model the dominant effect of precession on the orbital
phase. A later scheme for searching with precessing templates
was introduced in Ref.~\cite{Buonanno:2002fy}, which used unphysical
coefficients in the waveform expansion to model the effects of precession
in a parameter space of reduced dimension. The approach described in
Ref.~\cite{Buonanno:2002fy} was used to search for precessing signals in LIGO
data~\cite{Abbott:2007ai,Abbott:2009tt}. However, it was determined that
due to the increased response to the noise background, this
method did not offer any improvement in sensitivity
compared to a nonspinning search pipeline~\cite{VanDenBroeck:2009gd}.
The basic problem was that while the increased
parameter space offered an improvement in the signal power recovered for
precessing signals, the large, additional, unphysical parameter space
being searched greatly increased the rate of background triggers of the search.
Therefore, when estimating sensitivity at a constant rate of false alarm, the
precessing search loses sensitivity compared to a nonspinning or
aligned-spin search.

An alternative to these ``unphysical template families''
is to use a method that restricts to only physically possible template
waveforms. A method for doing this was first proposed in Ref.~\cite{Pan:2003qt}
and then further explored in Refs.~\cite{Buonanno:2004yd,Fazi:2009,Harry:2010fr}.
We will refer to this as the ``PTF'' (physical template family) approach. Our \ac{SMS} is motivated
from the PTF approach and we compare the two methods here.

The PTF approach considers single-spin systems and considers only the
$(\ell,|m|)=(2,2)$ modes in a corotating frame where the $z$ direction tracks the
orbital angular momentum. This is similar to our approach, except we do not restrict
ourselves to considering single-spin systems. The single-spin approximation was 
explored, however, in terms of an effective spin for double-spin systems, and
found to perform well, in Ref.~\cite{Buonanno:2004yd}.
The PTF approach parametrizes the single spin by the spin magnitude, $\chi_1$; the cosine of the angle between the spin and the
orbital angular momentum, $\kappa_1 = \mathbf{\hat{L}}_N \cdot \bm{\chi}_1$
and the azimuthal angle $\varphi$ between the projections of the spin
and the line of sight on the orbital plane, all these quantities being computed at some reference time.
The PTF SNR is then constructed by reexpressing $h_+$ and
$h_{\times}$ as the sum of five basis waveforms constructed using a spherical
harmonic ($\ell=2$) basis to express the waveform as~\cite{Pan:2003qt}
\begin{equation}
 h(t) = P_I \left[D, \theta, \psi, \phi, \iota, \varphi \right] Q^I \left[m_1, m_2, \chi_1, \kappa_1; \phi_c, t_c; t \right],
\end{equation}
where $I$ takes values $\in [1,5]$. Next, a maximization is performed over the five $P_I$ constants,
$\phi_c$ and $t_c$ to obtain the PTF SNR as described in Refs.~\cite{Buonanno:2004yd,Harry:2010fr}.

The statistical distribution of the PTF SNR, as described in Ref.~\cite{Harry:2010fr},
is template and noise curve
dependent. However, in the best case scenario, when the \ac{PDF} of the PTF SNR at high SNRs
takes the smallest values, the distribution of SNRs follows a $\chi^2$ distribution with 6
degrees of freedom. This can be compared to the distribution of values for the \ac{SMS},
as shown in Fig.~\ref{fig:statdetfigure}.
However, as summarized in Table~\ref{tab:intrinsicextrinsic}, the PTF SNR maximizes over more
parameters than the \ac{SMS}. Specifically, considering a single-spin template,
both the inclination angle and the azimuthal spin angle are maximized over in the PTF SNR,
but not maximized over in the \ac{SMS}.

The PTF SNR, as described above, is \emph{not} restricted to physical values of the
parameters. While the five $P^I$ values do depend on six physical parameters,
the values of $D, \theta, \psi, \phi$ are degenerate in a single detector,
and enter the waveform in only two combinations,
an overall amplitude scaling, and $F_+ / F_{\times}$. Therefore, only four independent
physical values can be measured from the $P^I$ and it then follows that
allowing a free maximization over five values allows
for unphysical combinations. The authors of Ref.~\cite{Buonanno:2004yd} explored this,
and described a numerical method for constraining the $P^I$ values to the
physical manifold. However, that method is computationally expensive, and was not
included in attempts to use the PTF approach as a search method~\cite{Fazi:2009,Harry:2010fr}.

In Ref.~\cite{Buonanno:2004yd} the authors considered searching for precessing
signals in data from only a single detector. Here we wish to consider the case
of a multidetector analysis. The standard approach is to measure the SNR
from data in each detector \emph{independently} and then look for
times where both detectors obtain a SNR over some threshold, within
some predefined time window and within some predefined window on the template
parameters that are gridded over.
No attempt is made to ensure consistency in the
parameters that are maximized over. For a two-detector aligned-spin search
one maximizes over time, an overall waveform phase, and an overall amplitude in
each detector. As long as the time difference is within that allowed due to the
light travel time between detectors it is always possible to find a physical
solution for the maximized parameters. For more than three detectors this is not the
case, and the aligned-spin coincident search will allow unphysical combinations of the
maximized parameters. In that case the computationally more expensive coherent
search~\cite{Bose:1999pj,Finn:2000hj, Pai:2000zt, Harry:2010fr,Macleod:2015jsa}
can offer some improvement in sensitivity by restricting the search to only
physically possible values of the maximized parameters.

In the PTF approach, even considering the constrained statistic, one measures
6 maximized parameters ($P^{1..4}$, $\phi_c$, $t_c$) in each detector.
With two detectors these are measured independently and so we obtain 12 independent
quantities measuring only 8
physical parameters. For the nonconstrained statistic 14 independent
quantities are maximized over, measuring the same 8 physical parameters.
This results in a significantly large
unphysical region of parameter space being searched over when considering
multiple detector searches and makes it
difficult to see sensitivity gains over the aligned-spin searches.

In the \ac{SMS}, we reduce six physical parameters to
four nondegenerate combinations of them that are maximized over.
For a single detector, the resulting values are
always consistent with some physical signal.
When filtering in two detectors, however, we have obtained eight 
independent measurements for only six physical quantities.
It therefore follows that, as in the PTF case, when filtering with
two or more detectors, some degree of unphysical freedom is still
allowed with this new statistic. However, this new statistic allows
\emph{less} unphysical freedom than in the PTF case and therefore
should offer a better chance to create a search that increases
sensitivity to precessing systems.

\begin{table*}[tbp]
\begin{tabular}{ c || c | c }
 Maximization scheme & Discrete parameters in the bank & Parameters continuously optimized over\\
  \hline
 Sky-maximized SNR & $m_1$, $m_2$, $\bm{\chi}_1$, $\bm{\chi}_2$, $\iota$ & $\phi_c$, $t_c$, $\theta$, $\phi$, $\psi$, $D$ \\
  PTF SNR (constrained max) & $m_1$, $m_2$, $\chi_1$, $\kappa_1$ & $\iota$, $\varphi$, $\phi_c$, $t_c$, $\theta$, $\phi$, $\psi$, $D$ \\
  \hline
\end{tabular}
\caption{\label{tab:intrinsicextrinsic}Comparison between the
  maximization schemes proposed in the PTF method~\cite{Pan:2003qt}
  and in this paper.  PTF lays a discrete bank in a 4d space and then
  ``continuously'' maximizes over the remaining $12-4=8$ parameters
  using a combination of analytic and numerical methods.  Note that
  the total number of parameters is 12 instead of 15 due to their
  restriction to single-spin systems. In this paper, we essentially
  follow the same route except that (i) we do not restrict ourselves
  to single-spin systems and (ii) we include $\iota$ and $\varphi$ in
  the discrete bank.  Note that in PTF the parameter $\varphi$
  (together with $\chi_1$ and $\kappa_1$) describes the orientation of
  the spin in the source frame. In our notation, these three
  parameters combine into $\bm{\chi}_1$.}
\end{table*}

\section{Template bank construction}
\label{sec:bank_cons}

We now apply the \ac{SMS} to generate banks of generic-spin
\ac{BBH} and \ac{NSBH} templates. We also quantify the expected
improvement in \ac{SNR} recovery from a simulated population of
precessing-spin systems when using these generic-spin template banks, compared to
aligned-spin-only banks. In each case, we first generate a bank of
aligned-spin templates, and then add precessing templates to the
aligned-spin bank in a second stage. By construction, since the
aligned-spin bank is a subset of the generic-spin bank, the
generic-spin template bank will outperform the aligned-spin bank
towards any putative signal when performance is measured in terms of
\ac{SNR} recovery. We remind the reader that comparing the expected
\ac{SNR} recovery between the two banks does not include the effect of
the increase in background event rate due to the increase in template
bank size and increased degrees of freedom incurred by using the
precessing templates. We consider this effect in the next section, in
which we demonstrate the application of the precessing template banks
in an analysis of simulated Advanced LIGO noise.

\subsection{Precessing bank construction method}
\label{ssec:stoch_bank_place}

The construction of efficient template banks to search for
nonspinning compact-binary mergers has been well explored in the
literature~\cite{Sathyaprakash:1991mt, Poisson:1995ef,
  Balasubramanian:1995bm, Owen:1995tm, Owen:1998dk, Babak:2006ty,
  Cokelaer:2007kx, Babak:2012zx}.  These methods define a metric in
the parameter space of the two masses, and use this metric to place a
hexagonal lattice in appropriate coordinates.  Recently, 
this method has been extended to place geometrical
lattices of aligned-spin waveform
templates~\cite{Brown:2012qf,Harry:2013tca}.  However, for the
parameter space of precessing compact-binary mergers, it is not clear
how to choose a coordinate system in which it is appropriate to lay a
lattice of waveform templates. Indeed, there is no reason to believe
that such an intrinsically flat parameter space exists for the
precessing parameter space.  In addition, current geometric methods
exclude the effects of merger and ringdown in the
waveform model. This can cause these banks to be suboptimal in the
\ac{BBH} parameter space where merger and ringdown are
important.

For these reasons, we make no attempt to employ a geometrical template
bank for our precessing search. Instead, we use a ``stochastic''
template bank construction
scheme~\cite{Babak:2008rb,Harry:2009ea,Ajith:2012mn}.  The stochastic
template scheme has the advantage that it is able to create a bank of
waveform signals for any parameter space, but the disadvantage that it
will require more templates to cover a given parameter space than a
geometrical lattice, and can be considerably more computationally
expensive~\cite{Harry:2009ea,Manca:2009xw}. In recent years, a number
of methods have been proposed to significantly speed up the generation
of stochastic banks~\cite{Ajith:2012mn,Fehrmann:2014cpa,Capano:2016dsf}.
We use a number of these methods here, along with
some new methods to optimize bank placement, which we describe in the
Appendix.

The general
stochastic approach works as follows. Begin with a seed bank $B$,
which may be empty. Then randomly choose the parameters of a putative
template signal and compute the corresponding template
waveform $g_\mathrm{prop}$. Then a match, $\mathcal{M}\left(g_\mathrm{prop}, h\right)$,
is computed
between $g_\mathrm{prop}$ and all templates $h \in B$, where
$\mathcal{M}\left(g_\mathrm{prop}, h\right)$ defines the
fraction of the signal power of $g_\mathrm{prop}$ that would be recovered
if using the template $h$ as a filter.
The fitting factor $\textrm{FF}$ is then defined as the match maximized over all
templates in the bank,
\begin{equation}
  \textrm{FF} \equiv \max_{h\in B} \mathcal{M}(g_\mathrm{prop},h).
\end{equation}
If the fitting factor falls below a given threshold---the
``minimal match''---then the proposed template is added to the bank
and $B'=B \cup\{h_\mathrm{prop}\}$ is set as the seed bank for the
next iteration. Otherwise, the proposed template is discarded, and
$B'=B$. The process repeats until a sufficiently high rejection rate
of proposed templates is achieved.

For aligned-spin placement, the match $\mathcal{M}(g_\mathrm{prop} ,h)$
defined from the phase-maximized matched-filter \ac{SNR} in Eq.~\eqref{eqn:max_likelihood_kphi} is
\bea
 \mathcal{M}(g_\mathrm{prop} ,h) &\equiv& \max_{\phi_c, t_c} \Re \left[\langle \hat{g}_\mathrm{prop} | \hat{h}(\phi_c, t_c)\rangle \right]
 \nonumber \\
 &=& \max_{t_c} \left| \langle \hat{g}_\mathrm{prop} | \hat{h}(0, t_c)\rangle \right|.
 \label{eq:aligned_spin_M}
\eea
We note that, as written, the maximization over time and phase shift is performed only on the
template $h$ and not on the proposed waveform $g_\mathrm{prop}$.
However, given
the assumptions that are used to construct Eq.~\eqref{eqn:max_likelihood_kphi}, a
phase or time shift in $h$ can be modeled by an opposite phase or time shift in $g_\mathrm{prop}$
and therefore the form given in Eq.~\eqref{eq:aligned_spin_M} serves to maximize over
a phase and time shift in both $h$ and $g_\mathrm{prop}$. Therefore, when choosing seed points for
aligned-spin stochastic template bank construction,
one only needs to choose the masses and spins.

For precessing waveforms, we define the match as
\begin{equation} \label{eq:precessing_spin_M}
 \mathcal{M}(h,g) \equiv \max_{\phi_c, t_c, u} \Re \left[\langle \hat{g}_{\mathrm{prop}} | \hat{h}(\phi_c, t_c, u)\rangle\right],
\end{equation}
where the maximization is performed as described in Eq.~\eqref{eqn:likelihood_phase_amp_u_maxed_003}.
In this case, a variation in the value of $u$ for $\hat{g}_\mathrm{prop}$ cannot be
written as a corresponding shift of the value of $u$ in $h$. Therefore,
in the precessing case, we maximize over $\phi_c$ and $u$ \emph{only}
in the template waveform. That is, when picking the putative template
signal we choose a sky location, construct $h(t)$ and compare that against
the $h_+$ and $h_{\cross}$ components of all templates in the bank, using
Eq.~\eqref{eqn:likelihood_phase_amp_u_maxed_003}. If a putative point is accepted
into the template bank, the sky location and $h(t)$ are discarded, and
only the $h_+$ and $h_{\cross}$ components of that point are retained.
Our choice not to maximize over $\phi_c$ and $u$ in $\hat{g}_\mathrm{prop}$
is taken to allow us to use Eq.~\eqref{eqn:likelihood_phase_amp_u_maxed_003}
directly when evaluating the fitting factor for potential precessing filter waveforms.
This choice will result in an increase in the number of proposal points needed to
construct a precessing template bank than if one were to consider also maximizing
over $\phi_c$ and $u$ in $\hat{g}_\mathrm{prop}$.
However, the final number of templates in the resulting template bank should
not be affected by this choice.

\subsection{Effective fitting factor}
\label{ssec:eff_fit_fac}

In order to quantify a template bank's performance in terms of \ac{SNR}
recovery, we use the notions of {\it signal recovery fraction} and {\it
  effective fitting factor} as figures of merit. These notions were defined
initially in Refs.~\cite{Buonanno:2002fy,Harry:2013tca}, and we redefine them here
for completeness.  Consider a template bank and a model $p(\bm{\upsilon})$ for
the distribution of source parameters.  We assume that sources are uniformly
distributed in volume, so that $p(\bm{\upsilon}) \propto r^2 p(\bm{\upsilon}')$
where $\bm{\upsilon}'$ denotes all parameters other than distance. For a
``perfect'' template bank, where all fitting factors are unity, the expected
total number of sources that would be observed above a \ac{SNR} threshold
$\rho_0$ is proportional to
\begin{equation}
 N_{\textrm{opt}} \propto \int \sigma^3(\bm{\upsilon}') p(\bm{\upsilon}') d\bm{\upsilon}',
\end{equation}
where $\sigma(\bm{\upsilon}')$ is the distance at which the expected \ac{SNR} to 
the signal with parameters $\bm{\upsilon}'$ is equal to $\rho_0$.
In reality, our template banks will not have a fitting factor of 1 for the
entire parameter space, and therefore the number of observed signals above a
\ac{SNR} threshold $\rho_0$ will be smaller than $N_{\textrm{opt}}$ according to
\begin{equation}
 N_{\textrm{obs}} \propto \int \textrm{FF}^3(\bm{\upsilon}') \sigma^3(\bm{\upsilon}') p(\bm{\upsilon}') \bm{\upsilon}',
\end{equation}
where $\textrm{FF}(\bm{\upsilon}')$ denotes the fitting factor
between the signal with parameters $\bm{\upsilon}'$ and the template bank.

We then define the ``signal recovery fraction'' $\alpha$ as the ratio between
$N_{\textrm{obs}}$ and $N_{\textrm{opt}},$
\begin{equation}
 \alpha \equiv \frac{N_{\textrm{obs}}}{N_{\textrm{opt}}} = 
      \frac{\int \textrm{FF}^3(\bm{\upsilon}') \sigma^3(\bm{\upsilon}') p(\bm{\upsilon}') d\bm{\upsilon}'}
           {\int \sigma^3(\bm{\upsilon}') p(\bm{\upsilon}') d\bm{\upsilon}'},
\end{equation}
which takes values between 0 and 1.   It is also convenient to express the
bank performance in terms of the ``effective fitting factor,'' defined as
\begin{equation} \label{eq:fitting_factor_111222}
  \textrm{FF}_\textrm{eff} \equiv \alpha^{1/3},
\end{equation}
which can be interpreted as the average SNR recovered for the observed
population of sources. Including the $\sigma$ factors means that signals that
would not be seen at a large distance have only a weak effect on the signal
recovery fraction. This includes signals that are poorly aligned with respect to
the detector, and signals that have intrinsically low \ac{GW} luminosity. In
contrast, favorably oriented, intrinsically bright sources will have the largest
effect on this measure.

We estimate the signal recovery fraction numerically by Monte Carlo integration,
choosing a random set of source parameters $S=\{\bm{\upsilon}_i'\}_{i=1}^N$
according to a sampling distribution $q(\bm{\upsilon}')$. The signal recovery
fraction is then given by
\begin{equation}
  \alpha \approx \frac{\sum_{i=1}^N \textrm{FF}^3_i \sigma^3_{i} (p_i/q_i) }{\sum_{i=1}^N  \sigma^3_i (p_i/q_i)},
\end{equation}
where $p_i/q_i=p(\bm{\upsilon}'_i)/q(\bm{\upsilon}'_i)$ corrects the sampling
distribution $q$ to match the desired astrophysical distribution $p$.
Table~\ref{tbl:mass_spin_params} summarizes the distribution
$q(\bm{\upsilon}')$ that we use when drawing signals to evaluate both our NSBH and \ac{BBH} template
banks. The template banks are constructed using the same limits on physical parameters\footnote{
As discussed in Sec.~\ref{sec:waveforms}, the frequency-domain NSBH precessing
waveform approximant that is used when placing the NSBH precessing template bank is a single spin model.
Therefore, NSBH precessing templates all have $\chi_2=0$, although the aligned-spin NSBH template bank contains
templates with $\chi_2 \neq 0$. When calculating the effective fitting factor and associated quantities all NSBH
templates and simulated signals are modeled using the double-spin time-domain TaylorT2 waveform model.}.

We wish to sample well all points in our parameter space, and so we choose
sources uniformly distributed in component masses within the bounds for the
\ac{BBH} and NSBH parameter spaces given in Table~\ref{tbl:mass_spin_params}.
However, such a choice leads to an effective fitting factor that is dominated by
high mass systems since, to leading order, $\sigma \propto
\mathcal{M}^{5/6}$. This would be the correct figure of merit if the
distribution of masses for compact-binary mergers in the Universe was uniform in
component masses. However, this distribution is not well
understood~\cite{Ozel:2010su,Farr:2010tu}. To obtain a figure of merit for our
banks that more evenly averages over the mass space, we therefore correct the
sampling distribution by a factor $p/q=\mathcal{M}^{-5/6}$ to approximate a prior in
mass such that the \emph{observation rate} does not change as a function of
component masses. Thus, we evaluate the signal recovery fraction from our
simulations by the formula
\begin{equation}
 \label{eq:eff_fit_fac_1}
 \alpha \approx \left(\frac{\sum_{i=1}^N \textrm{FF}^3_i \sigma^3_{i} \mathcal{M}_i^{-5/6}}
          {\sum_{i=1}^N \sigma^3_i \mathcal{M}_i^{-5/6}} \right).
\end{equation}
To minimize issues of uncertainty in the real astrophysical mass distribution,
we also report below results restricted to relatively small mass bins. However,
when reporting results in bins of spin, the chosen mass distribution will
matter, and this distribution more evenly weights the simulations.

\subsection{Effectualness of precessing template banks}
\label{ssec:testing_banks}

\begin{table}[tbp]
  \begin{tabular}{c|c|c|c|c}
    \parbox[t]{1.5cm}{Parameter \\ space} & \parbox[t]{1.2cm}{Minimal \\ match} & Spin & Templates & Eff. FF\\ \hline
    \multirow{2}{*}{NSBH} & 0.97 & Aligned & 146,315 & 0.948 \\
& 0.90 & Precessing & 1,583,079 & 0.976 \\
    \multirow{2}{*}{\ac{BBH}} & 0.97 & Aligned & 23,948 & 0.984 \\
& 0.90 & Precessing & 237,909 & 0.988 \\
  \end{tabular}
    \caption{\label{tbl:tmpltbank_sizes}Sizes and effective fitting factors for the \ac{BBH}
    and NSBH template banks. These cover the same range of parameters from
    which the corresponding signal set is drawn from, as described in
    Table~\ref{tbl:mass_spin_params}. When computing matches and constructing the template
    banks a lower frequency cutoff of 30Hz is used and the Advanced LIGO ``early''
    noise curve prediction~\cite{Aasi:2013wya}. The effective 
    fitting factor reported in the last column is computed using Eq.~\eqref{eq:fitting_factor_111222}.}
\end{table}

\begin{figure}[tb]
  \includegraphics[width=\columnwidth]{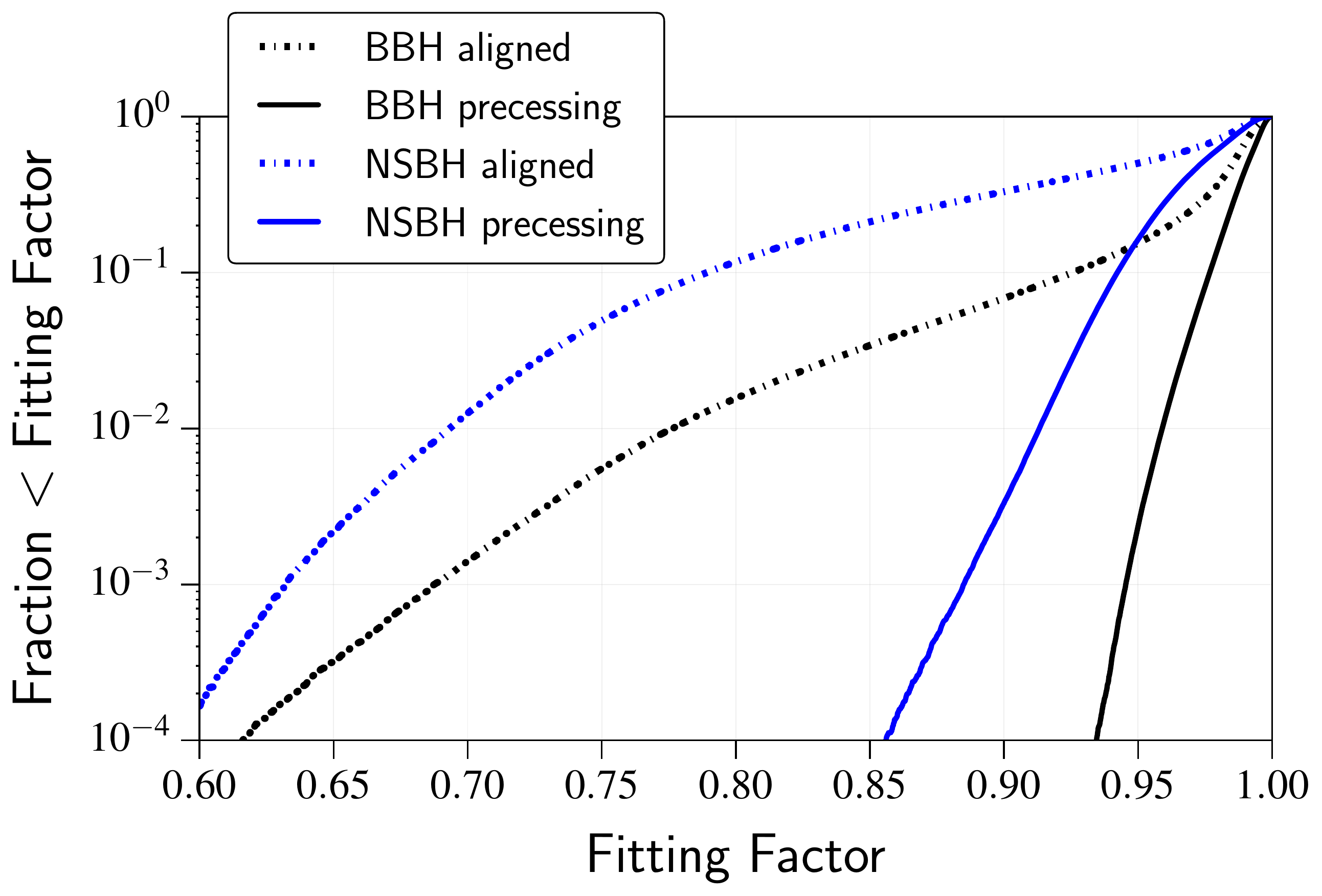}
  \caption{\label{fig:fitting_factor_dist}The distribution of fitting factors for the aligned-spin
  and precessing-spin template banks summarized in Table~\ref{tbl:tmpltbank_sizes} and
  covering the parameter spaces defined in Table~\ref{tbl:mass_spin_params}.}
\end{figure}

We now
generate template banks to cover the parameter spaces that are summarized
in Table~\ref{tbl:mass_spin_params}. Our aligned-spin banks, which form
the seed for the precessing banks, are generated using a minimal match
of 0.97, which matches the value used for aligned-spin
searches of Advanced LIGO and Advanced Virgo data~\cite{Capano:2016dsf}.
The precessing banks become unmanageably large
with our current methods if generated with a minimal match of 0.97.
We therefore use a minimal match of 0.9 when completing our
aligned-spin banks with precessing waveforms.

The sizes of the resulting template banks and the corresponding
effective fitting factors of those banks are summarized in
Table~\ref{tbl:tmpltbank_sizes}, and the distribution of fitting
factors for each bank is plotted in
Fig.~\ref{fig:fitting_factor_dist}. We find that the precessing
template banks are roughly an order of magnitude bigger than their
aligned-spin counterparts, and that, on average, aligned-spin banks
are already performing reasonably well when searching for precessing
systems.  The \ac{BBH} aligned-spin bank is more effective at
recovering precessing \ac{BBH} signals
($\mathrm{FF}_\mathrm{eff}=0.984$) than the NSBH aligned-spin bank is
at recovering precessing NSBH systems
($\mathrm{FF}_\mathrm{eff}=0.948$).  However, both aligned-spin bank
distributions show long tails of precessing systems that are recovered
with fitting factors less than 0.90.  When using precessing template
banks, these tails are significantly reduced. The effective fitting
factor also increases to $\mathrm{FF}_\mathrm{eff}=0.988$ for the
\ac{BBH} parameter space and $\mathrm{FF}_\mathrm{eff}=0.976$ for the
NSBH parameter space. While the increase in the overall effective
fitting factor---averaged over the full parameter space---is small
our precessing template banks seem to reduce an observational bias
against highly precessing signals that will be present in current
aligned-spin searches. We explore this further in the next section
when we put these numbers into context by taking into account the
increase in background incurred from filtering the data against a
larger number of templates.

\section{Assessing the sensitivity of the precessing search}
\label{sec:searches}

The increase in templates from including the effects of precession,
coupled with the fact that the precessing-spin templates on average
produce larger background \ac{SNR} values than the aligned-spin
templates with the \ac{SMS}, leads to an increase in the rate of false
alarms at a given \ac{SNR}.  Correctly estimating the increase in
observed signals with our precessing template banks requires assessing
the sensitivity of the aligned-spin and precessing searches at a
constant false-alarm rate, which we take on in this section. To do so,
we incorporate the \ac{SMS} into the PyCBC search pipeline described
in \cite{Canton:2014ena,Usman:2015kfa} and, using each of the banks
constructed in the previous section, perform a two-detector analysis
of Gaussian noise. These analyses give us a direct measurement of the
increase in background trigger rate, which we then combine with the
fitting factor calculations above to estimate the change in detection
rate of compact-binary systems at fixed false-alarm rate when using
precessing templates instead of only aligned-spin templates.

\subsection{Mapping between signal-to-noise ratio and false-alarm rate}
\label{ssec:searches_background}

\begin{figure}[t]
  \includegraphics[width=\columnwidth]{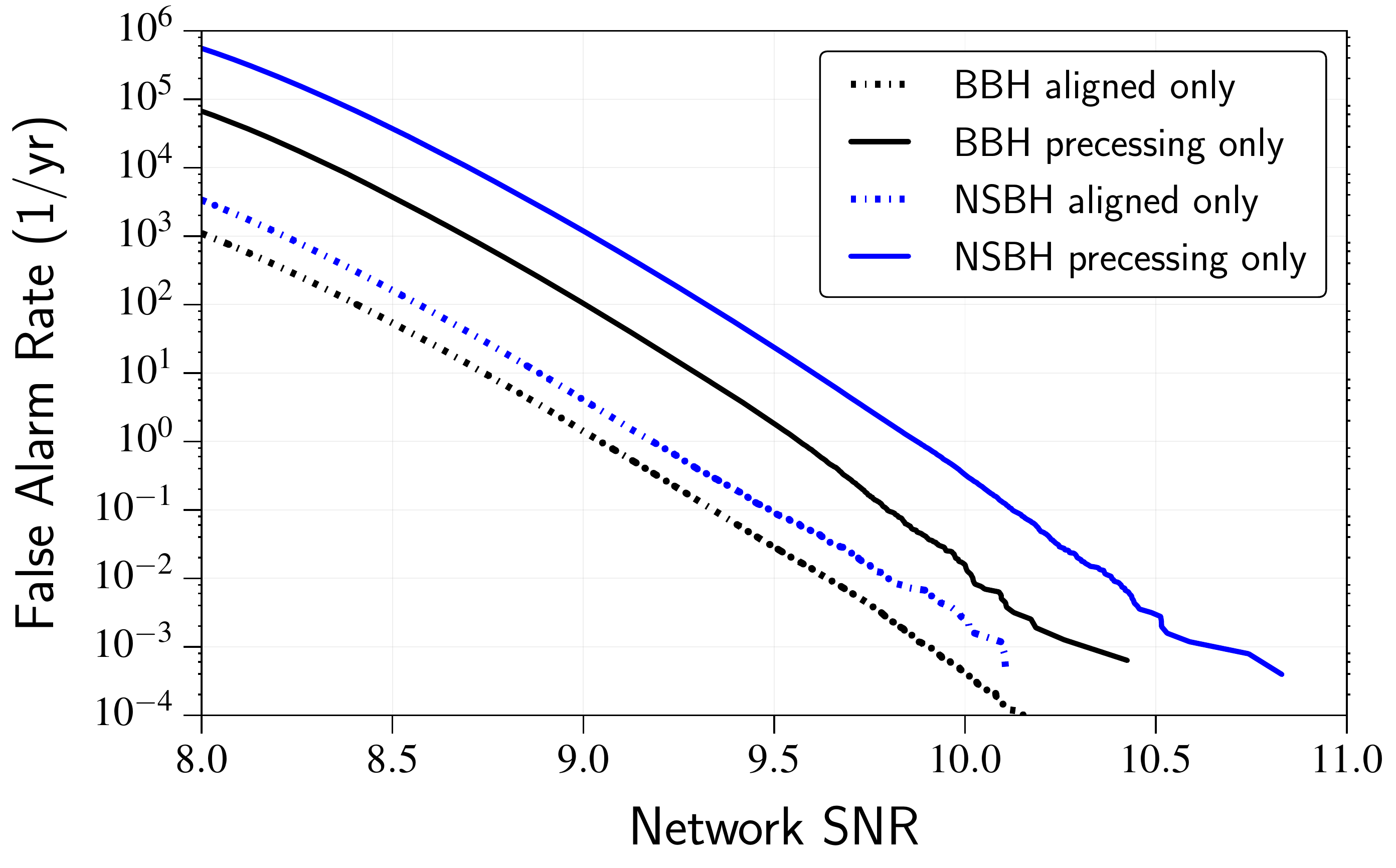}
  \caption{\label{fig:analysis_thresholds} The rate of coincident
    background events as a function of the network SNR $(\rho_H^2 +
    \rho_L^2)^{1/2}$ plotted for the \ac{BBH} and \ac{NSBH}
    aligned-spin-only and precessing-spin-only analyses. The full
    precessing search background is obtained by combining the
    aligned-spin-only and precessing-spin-only analyses, as described
    in the text.}
\end{figure}

\begin{table}[tb]
  \begin{tabular}{c|c|c}
    \parbox[t]{2.5cm}{\hfill \break Bank \break } & \parbox[t]{2.5cm}{SNR threshold at false-alarm rate of $10^{-2} yr^{-1}$} &
           \parbox[t]{2.5cm}{SNR threshold at false-alarm rate of $0.5 \times 10^{-2} yr^{-1}$} \\ \hline
    BBH aligned & 9.64 & 9.73 \\
    BBH precessing & Not applicable & 10.10\\
    NSBH aligned & 9.79 & 9.92  \\
    NSBH precessing & Not applicable & 10.44 \\
  \end{tabular}
  \caption{\label{tbl:bkg_thresholds}SNR thresholds at the false-alarm rate values used in this study
  for our BBH and NSBH aligned-spin-only and precessing-spin-only template banks.}
\end{table}

In performing the precessing search, we split the original
aligned-spin templates from the templates that were added to cover the
precessing space and analyze these two sets separately. The
``precessing-spin search'' is then formed by combining the
aligned-spin-only analysis with the precessing-spin-only analysis,
giving equal weight to each ``subsearch.'' This choice amounts to
assuming that an event is equally likely to appear in the aligned-spin
set of templates as in the precessing set. A better choice for this
could be made, but other studies suggest it will not drastically
affect our conclusions~\cite{Prix:2009tq,Dent:2013cva}; we leave this
to future work. When assessing the sensitivity of the aligned-spin
search alone, we use a false-alarm threshold of 1 in 100 years. When
assessing the sensitivity of the combined search, we use a false-alarm
threshold of 0.5 in 100 years for \emph{both} the aligned-spin-only
and precessing-spin-only subsearches, corresponding to a false-alarm
threshold of 1 in 100 years for the full precessing-spin search.

The result of these analyses on the aligned-spin-only and
precessing-spin-only template banks covering the \ac{BBH} and
\ac{NSBH} parameter spaces can be seen in
Fig.~\ref{fig:analysis_thresholds}. The relevant thresholds are
enumerated in Table~\ref{tbl:bkg_thresholds}. The table shows that a
signal that appears in the BBH aligned-spin bin with a SNR of 9.64
would be deemed as significant in the BBH aligned-spin search as a
signal that appears in the BBH precessing-spin bin with a SNR of
10.28 in the full BBH precessing-spin search. Thus, a precessing BBH
signal would only be found with higher significance by the
precessing-spin search if the precessing-spin templates increase its
SNR by at least the factor $10.28/9.64 \approx 1.07$. Similarly, a
precessing NSBH signal would only be found with higher significance by
the NSBH precessing-spin search if the precessing-spin templates
increase its SNR by at least the factor $10.44/9.79 \approx 1.07$. On
the other hand, a given aligned-spin signal will always be found with
a lower significance by the precessing-spin search, corresponding to a
loss of volume to aligned-spin systems of about $(9.73/9.64)^3 \approx
1.03$ for BBH and $(9.92/9.79)^3 \approx 1.04$ for NSBH. Thus, the
precessing search approach proposed here stands to lose at most $4\%$
of signals that are already well covered by aligned-spin templates,
but will improve the detection of any system whose fitting factor
improves by more than 7\% when going from the aligned-spin bank to the
precessing-spin bank.

\subsection{Comparing sensitivity methodology}
\label{ssec:searches_sensitivity}

The signal recovery fraction and effective fitting factor defined in Sec.~\ref{ssec:eff_fit_fac}
provide a useful measure of the fitting factor averaged
across the parameter space. However, these measures do not take into account the background increase as we
discussed in the section above. To do this, we replace the $\sigma_i$ factor in the
numerator of Eq.~\eqref{eq:eff_fit_fac_1}---which we take here to denote the distance at which
a simulation will be recovered with a \ac{SNR} of 1, equivalent to setting $\rho_0=1$ in the notation
of Sec.~\ref{ssec:eff_fit_fac}---with $\sigma_i/ \rho_{\mathrm{thresh}}$, where
$\rho_{\mathrm{thresh}}$ is the threshold taken from Table~\ref{tbl:bkg_thresholds}. This
gives the distance at which a simulation will be recovered with a \ac{SNR} of
$\rho_{\mathrm{thresh}}$. For the aligned-spin search, the threshold-dependent signal recovery fraction is then written as
\begin{equation}
 \alpha_{\mathrm{aligned}} =
 (\rho_{\mathrm{thresh}})^{-3} \alpha
\end{equation}
For the precessing-spin search, which consists of the aligned-spin-only
and precessing-spin-only subsearches, the weighting factor
depends on the two subsearches according to
\begin{equation}
 \alpha_{\mathrm{combined}} = \left( 
 \frac{ \sum_{i=1}^N (\mathrm{FF_{weighted}})_i \, \sigma_i^3
        \mathcal{M}_i^{-5/6} }{\sum_{i=1}^N
\sigma_i^3 \mathcal{M}_i^{-5/6} } \right),
\end{equation}
where we define
\begin{equation}
 (\mathrm{FF_{weighted}})_i = \max_{j} \left\{(\rho_{\mathrm{thresh}})_j^{-3} \textrm{FF}_{i,j}^3\right\},
\end{equation}
and the index $j$ runs over the two subsearches.
This measure can be used to compute the \emph{relative} sensitivities
between search configurations. For the searches we describe above, the relative search sensitivity is computed as
\begin{equation}
\label{eq:defrelsearchsensitivity}
 \beta = \frac{\alpha_{\mathrm{combined}}}{\alpha_{\mathrm{aligned}}}.
\end{equation}
Here $\alpha_{\mathrm{aligned}}$ is computed using the aligned-spin bank results and the threshold
from the aligned-spin bank at a false-alarm rate of $10^{-2}$. $\alpha_{\mathrm{combined}}$ is
computed using both the precessing and aligned-spin results taking the thresholds for both banks
respectively at a false-alarm rate of $0.5 \times 10^{-2}$.

\subsection{Results: BBH parameter space}
\label{ssec:resultsBBH}

When we include the varying \ac{SNR} threshold, as discussed above, and evaluate the
relative search sensitivity between the aligned-spin template bank and the generic-spin
template bank as defined in Eq.~\eqref{eq:defrelsearchsensitivity}
we find $\beta = 0.978$. This means that, given our assumed distribution of signals (reweighed by chirp mass), our combined
search is on average slightly less sensitive than the aligned-spin search as far as the total number
of detections is concerned in our \ac{BBH} parameter space.
However, this precessing search could still allow us to recover signals
in specific ``highly precessing'' regions of parameter space that would not be observed with the
aligned-spin search.
As a first step in addressing this question, we need to understand which, if any, regions of parameter space are
sufficiently precessing that we are gaining sensitivity when using our precessing search method.

\begin{figure*}[tb]
  \includegraphics[width=\columnwidth]{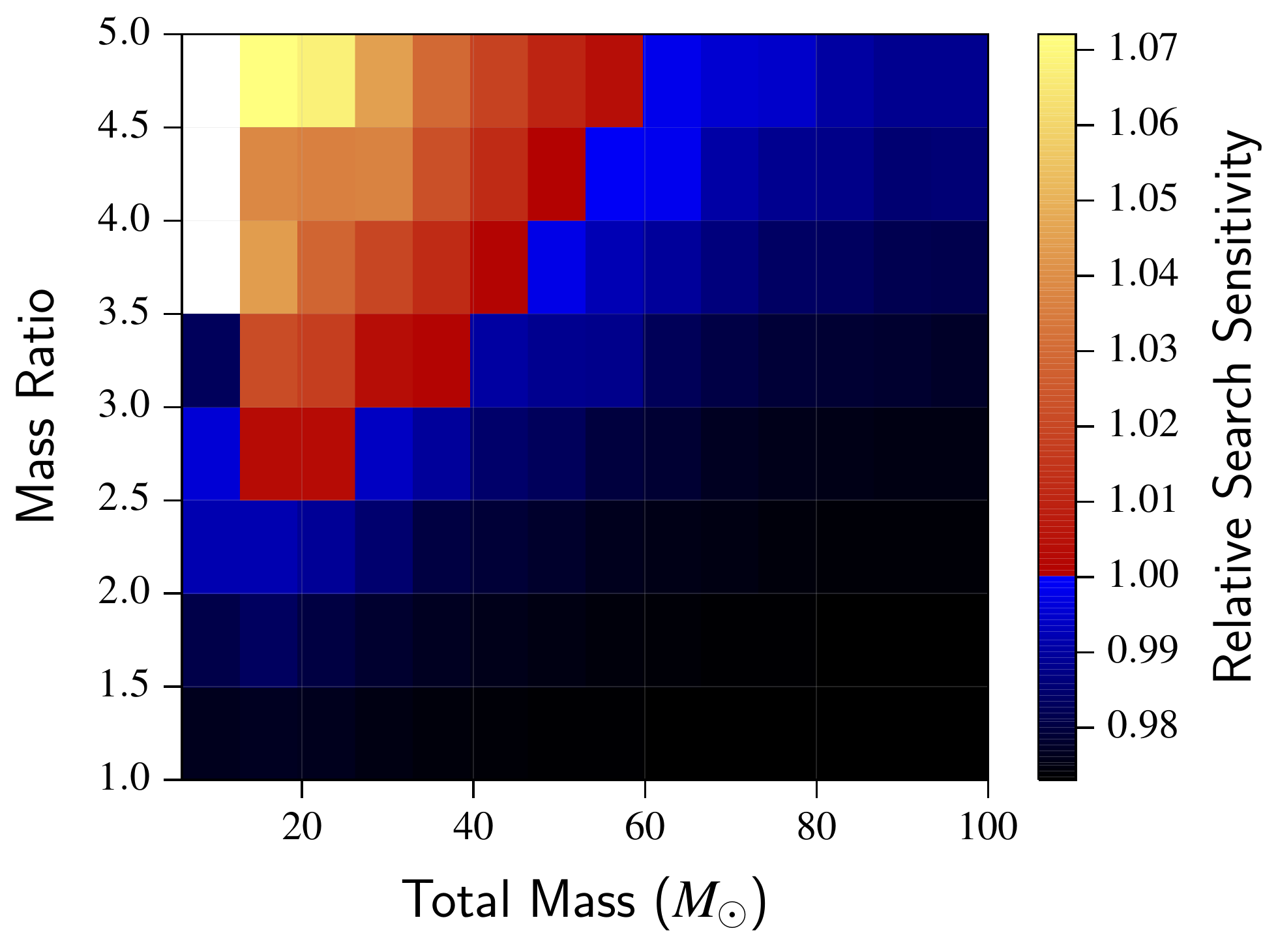}
  \includegraphics[width=\columnwidth]{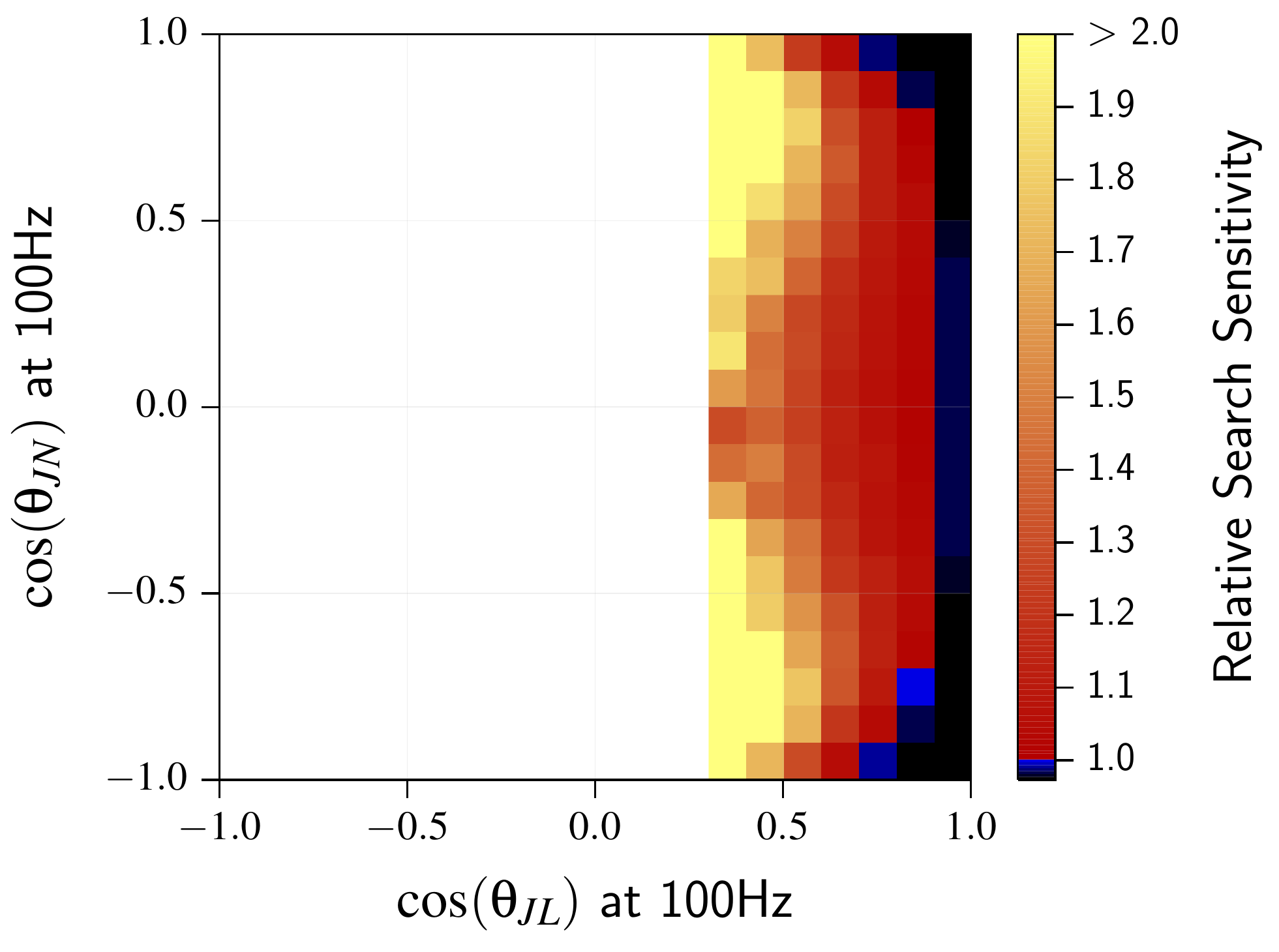}
  \caption{\label{fig:BBHsensitivity}The relative increase in sensitivity from performing an aligned-spin \emph{and}
  precessing search compared to performing only the aligned-spin search is plotted 
  in the (total mass, mass ratio) space in the left panel and as a function of opening angle and inclination
  (defined in the text) in the right panel. For BBH systems with the mass range considered here, the orbital
  angular momentum $\mathbf{L}$ is always larger than the component angular momenta $\mathbf{S}_i$ and therefore
  the angle between $\mathbf{L}$ and $\mathbf{J}$ can only take a restricted set of values, as can be seen in
  the right panel of this figure.}
\end{figure*}

Visualizing any quantity in the precessing parameter space is complicated by the large number
of dimensions and one has to resort to choosing two dimensional slices and marginalizing over the
remaining parameters. One slice that is traditionally shown considers the component masses, or equivalently
the total mass and the mass ratio of the system (left panel of Fig. \ref{fig:BBHsensitivity}). For close to
equal mass binaries, where precessional effects are expected to be small, the combined search leads to a
loss in sensitivity (always smaller than $3\%$). Only for mass ratios close to the upper boundary of our
parameter space do we obtain some mild sensitivity improvement of up to $7\%$.

The increase in sensitivity of the precessing search as the mass ratio increases is an expected consequence
of the fact that the magnitude of the orbital angular momentum $\mathbf{L}$ decreases as mass ratio increases,
thereby allowing larger opening angles between $\mathbf{L}$ and
$\mathbf{J}=\mathbf{L}+\mathbf{S}_1+\mathbf{S}_2$. Larger mass ratios also imply more precessional cycles in
band which contributes to our precessing search performing better. A similar effect is also obtained by lowering
the total mass, which explains why the sensitivity improves to the left of the plot.

\begin{figure*}[tb]
  \includegraphics[width=\columnwidth]{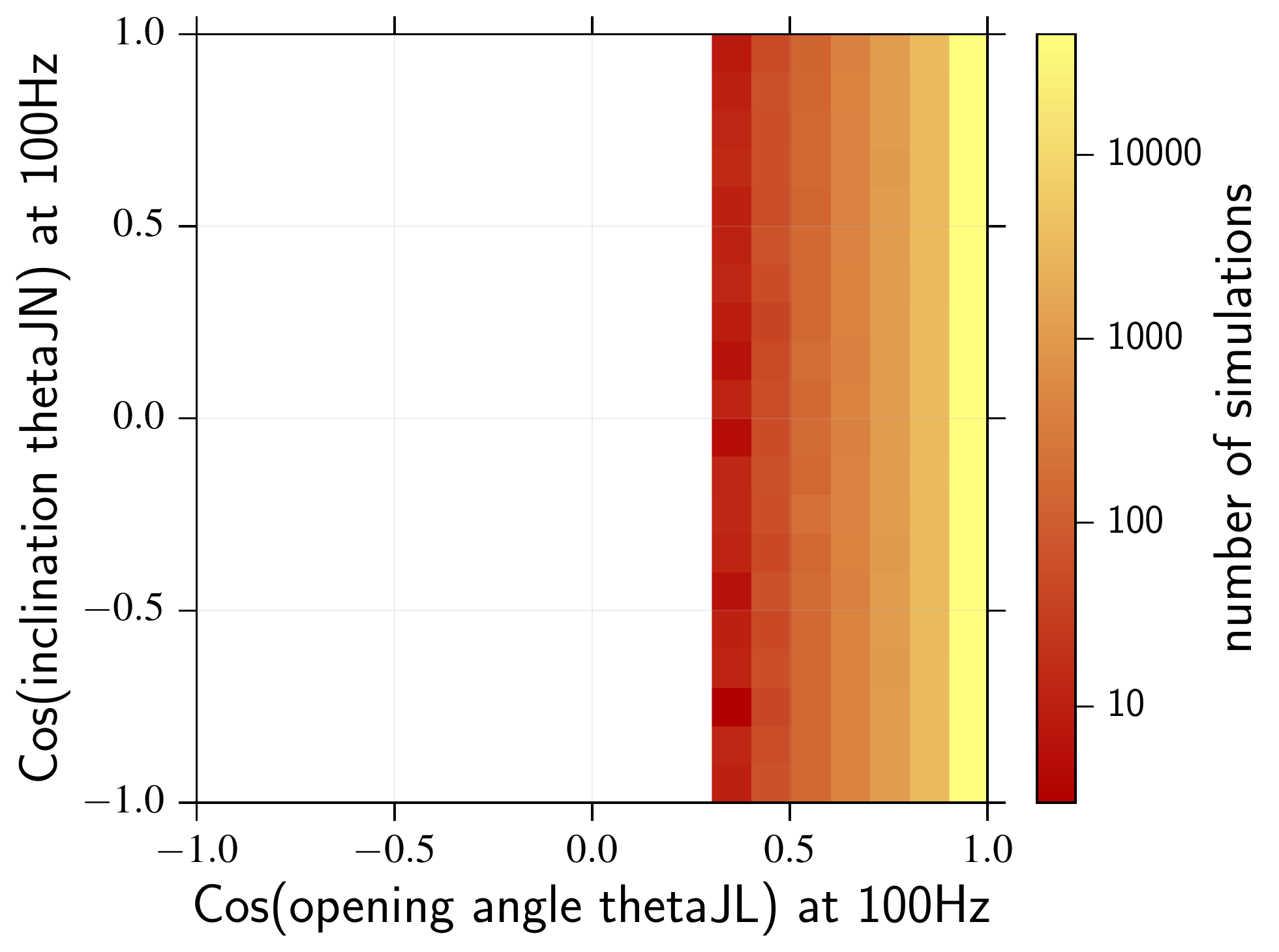}
  \includegraphics[width=\columnwidth]{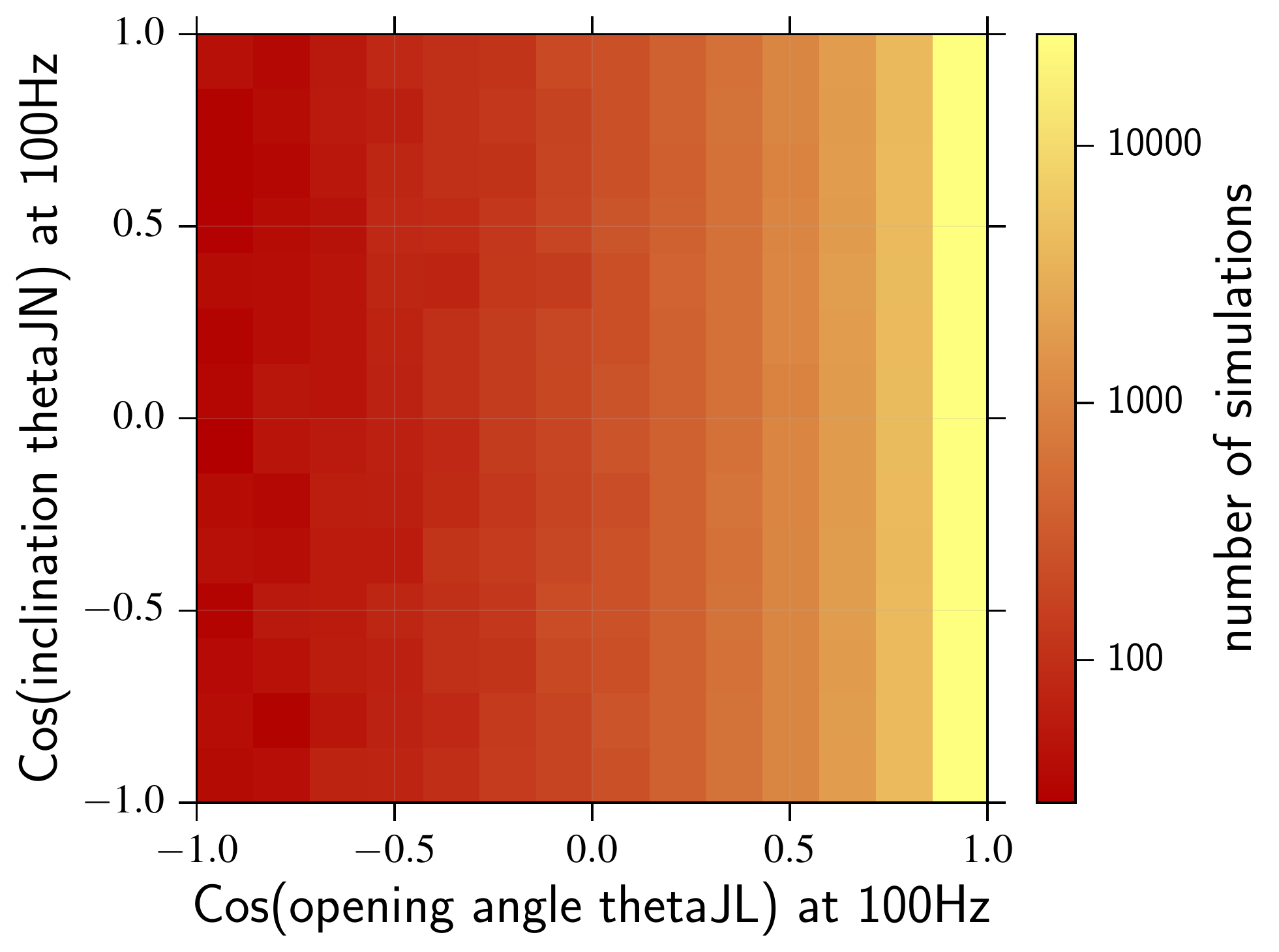}
  \caption{\label{fig:numsims}The number of simulations performed as a function of opening angle and inclination
  (defined in the text) for our BBH (left) and NSBH (right) parameter spaces.
  For BBH systems with the mass range considered here, the orbital
  angular momentum $\mathbf{L}$ is always larger than the component angular momenta $\mathbf{S}_i$ and therefore
  the angle between $\mathbf{L}$ and $\mathbf{J}$ can only take a restricted set of values, as can be seen in
  the left panel of this figure.
  }
\end{figure*}

A more appropriate set of two variables to identify highly precessing regions has been used 
in the right panel of Fig.~\ref{fig:BBHsensitivity}: the opening
angle between $\mathbf{J}$ and $\mathbf{L}$ introduced above and the inclination with which the
system is seen by the observer, defined as the angle between $\mathbf{J}$ and the line of sight
$\mathbf{N}$. The first quantity can be thought of as characterizing the intrinsic amount of
precession in the system. The inclination modulates how much precession an observer would see~\cite{Brown:2012qf}.
Of course, both angles actually
vary during the coalescence of the binary---the inclination varies on the radiation reaction time scale whereas
the opening angle can have modulations on the precessional time scale on top of the secular evolution
on the radiation reaction time scale---and therefore the angles must be evaluated at some reference
frequency. Here, we choose to use 100Hz as it roughly corresponds to the peak sensitivity of the
predicted Advanced LIGO noise curve~\cite{Aasi:2013wya}. We find regions where the sensitivity 
increases by a factor of larger than 2 with respect to the aligned-spin search. However, we note that it is 
extremely rare for signals to lie in these regions given the simulation distribution we are using. We
illustrate this in Fig.~\ref{fig:numsims} where we show the number of simulations as a function
of these spin angles.

\begin{figure*}[tb]
  \includegraphics[width=\columnwidth]{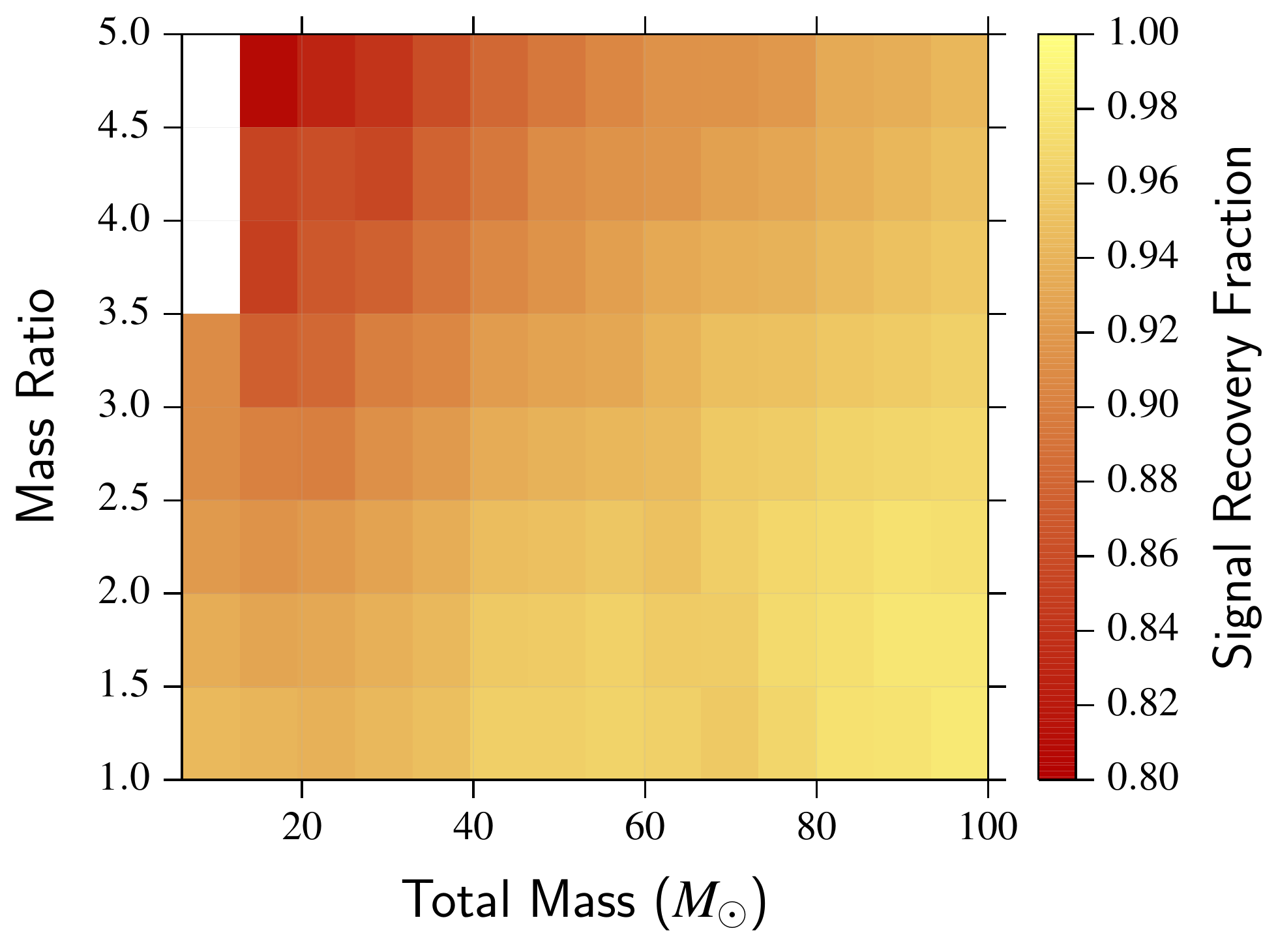}
  \includegraphics[width=\columnwidth]{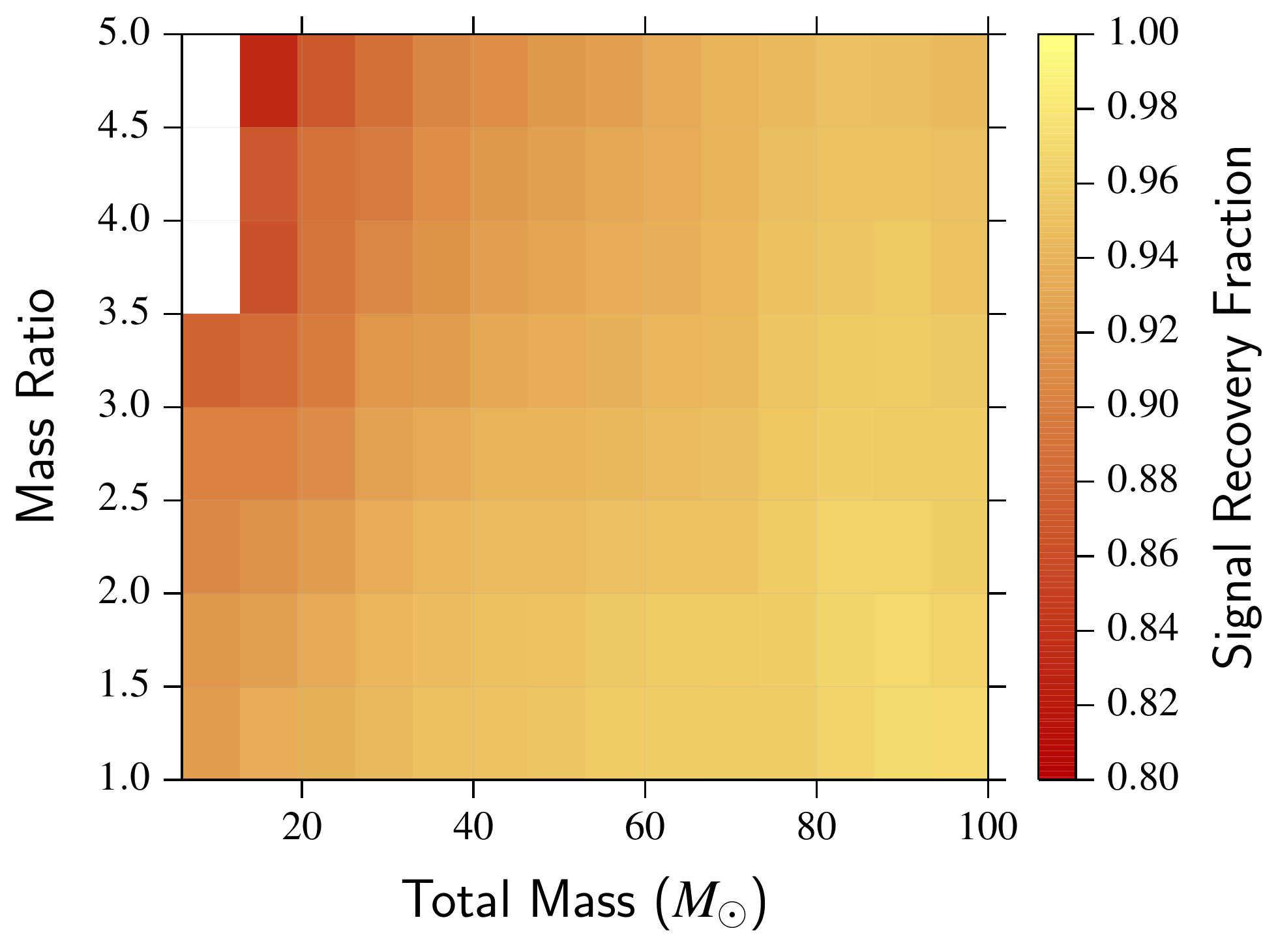}
  \caption{\label{fig:BBHWaveformCompare}Signal recovery fraction as a function of total mass and mass ratio
  comparing different waveform approximants. The left panel shows the signal recovery fraction of our set of precessing
  IMRPhenom simulations when using our aligned-spin IMRPhenom template bank. The right panel shows the signal recovery fraction
  of a set of precessing EOBNR simulations, with the same parameter distribution as the IMRPhenom simulations, recovered
  with an aligned-spin EOBNR template bank constructed as described in Ref.~\cite{Capano:2016dsf}.
  }
\end{figure*}

The results in this section are dependent on the waveform model used. However,
we expect that this dependence is only a weak one, and our results are still valid when
using template banks constructed for other waveform approximants. To check this, we
repeat our results using the EOBNR approximant, introduced earlier in Sec.~\ref{sec:waveforms}.
Unfortunately, at the current time, producing a precessing template bank using the EOBNR
approximant is not possible due to the time necessary to generate precessing EOBNR waveforms.
However, as described in Ref.~\cite{Capano:2016dsf} we can generate an aligned-spin EOBNR
template bank. In Fig.~\ref{fig:BBHWaveformCompare} we compare the ability to recover
precessing EOBNR waveforms using an aligned-spin EOBNR template bank,
with our ability to recover IMRPhenom signals using our aligned-spin IMRPhenom template bank.
We can see that the two panels in the figure
are largely indistinguishable.

\subsection{Results: NSBH parameter space}
\label{ssec:resultsNSBH}

We now consider the NSBH parameter space,
defined in Table~\ref{tbl:mass_spin_params}. When including the varying \ac{SNR} threshold, using the values in
Table~\ref{tbl:bkg_thresholds}, and Eq.~\eqref{eq:defrelsearchsensitivity} we find the relative
search sensitivity, evaluated at constant false-alarm rate, is $\beta = 1.014$. This means that we expect
to recover 1.4\% more signals averaged across the NSBH parameter
space when using the precessing search compared to the aligned-spin search. This number does,
of course, depend on the distribution of parameters that we chose---in
Table~\ref{tbl:mass_spin_params}---for our NSBH space.

\begin{figure*}[tb]
  \includegraphics[width=\columnwidth]{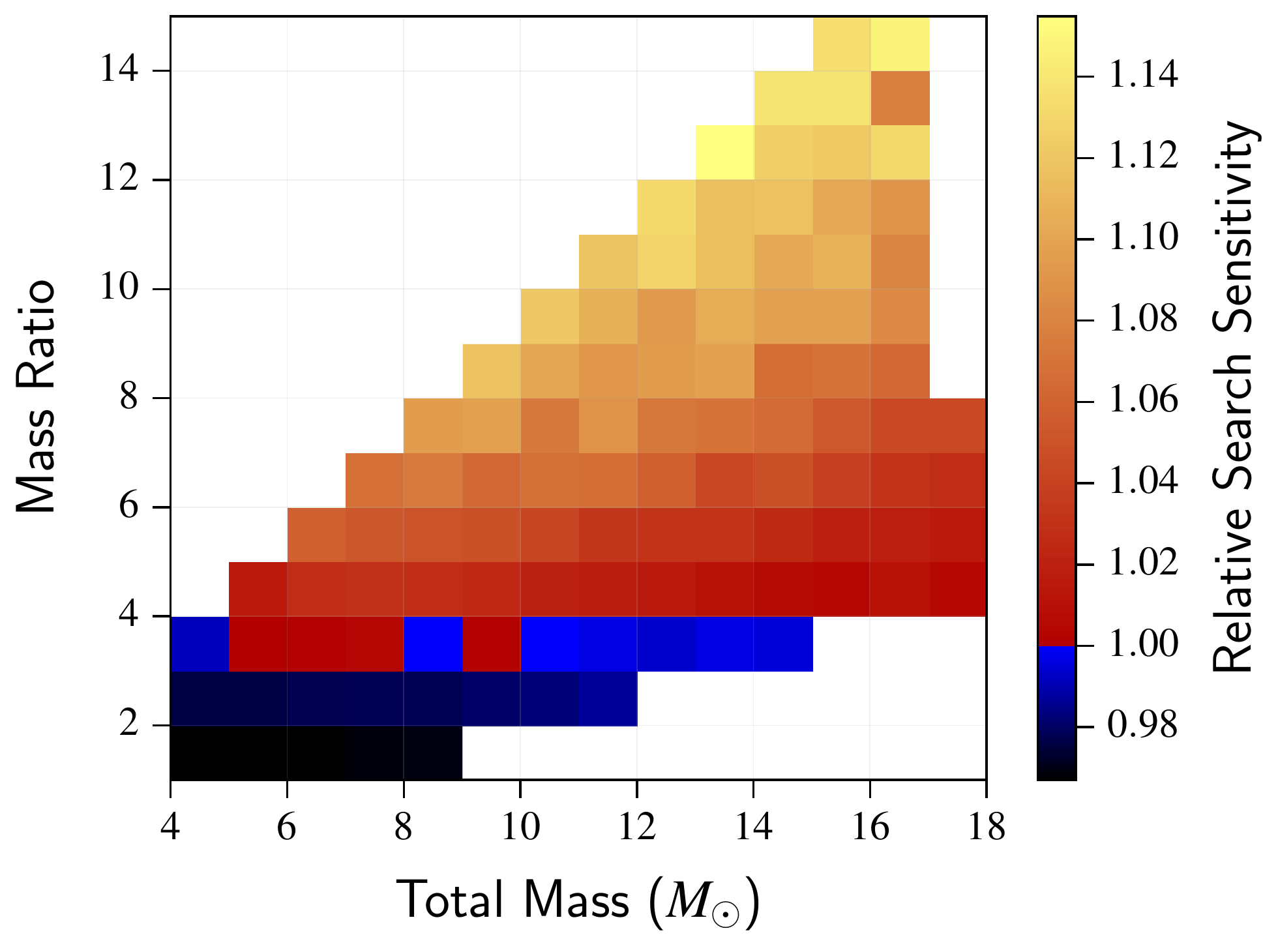}
  \includegraphics[width=\columnwidth]{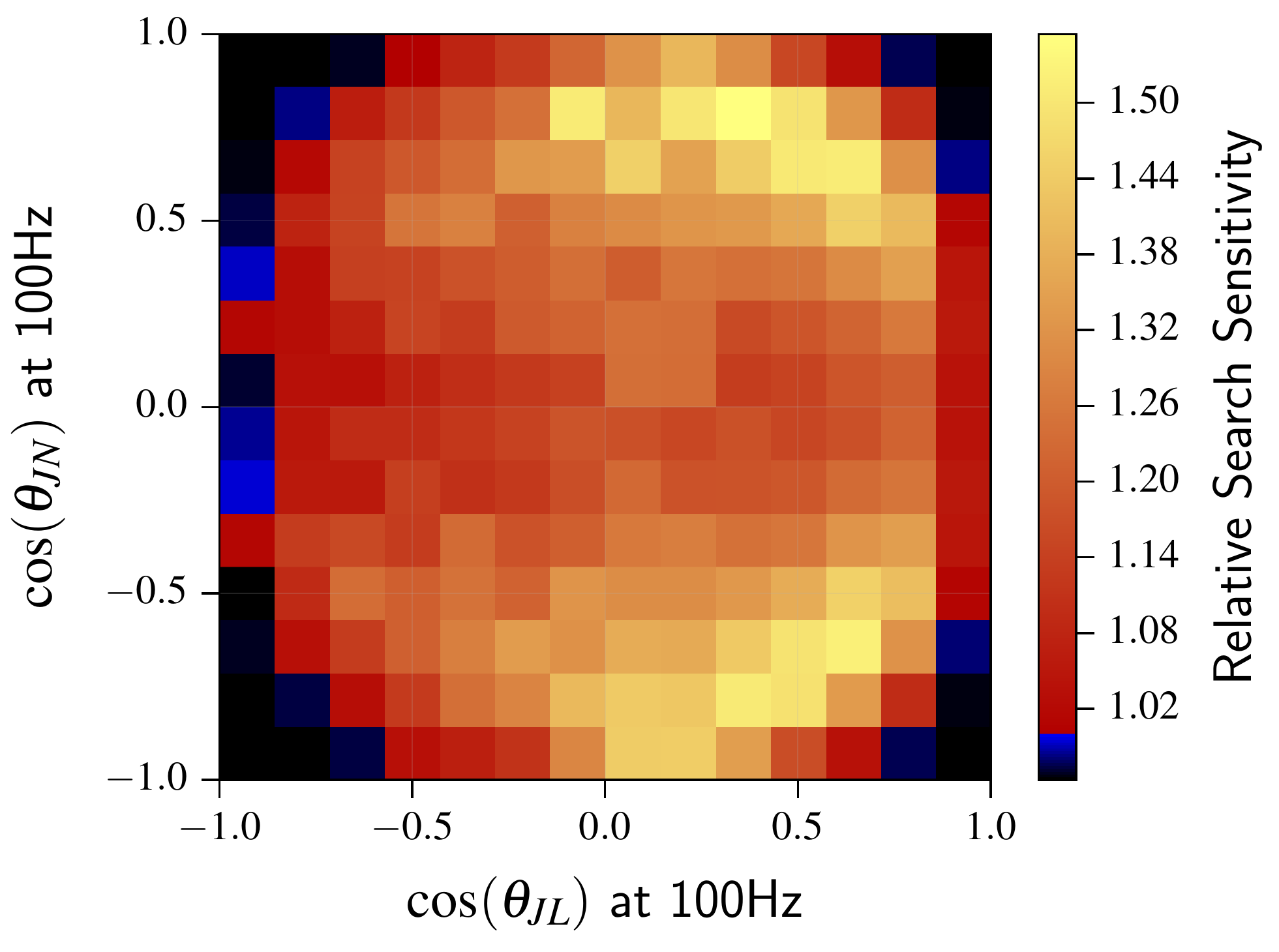}
  \caption{\label{fig:earlynsbh_analysis_results2}The relative increase in sensitivity from performing an aligned-spin \emph{and}
  precessing search compared to performing only the aligned-spin search is plotted 
  in the (total mass, mass ratio) space in the left panel and as a function of opening angle and inclination
  (defined in the text) in the right panel.}
\end{figure*}

In Fig.~\ref{fig:earlynsbh_analysis_results2}, we show the relative search sensitivity as a function
of the mass ratio and total mass (left) and also the angles between $\mathbf{J}$ and $\mathbf{N}$
and between $\mathbf{J}$ and $\mathbf{L}$ (right).
The relative search sensitivity as a function of total mass and mass ratio
(left panel of Fig.~\ref{fig:earlynsbh_analysis_results2}) shows similar trends to the corresponding ones
for \ac{BBH}, but given mass ratio values up to 15, we see larger relative search sensitivities, up to $1.14$ at
a mass ratio of 15. On the right panels of Figs.~\ref{fig:earlynsbh_analysis_results2} and \ref{fig:numsims}
we notice that systems in our NSBH parameter space are able
to cover all values of the angle between $\mathbf{J}$ and $\mathbf{L}$, which was not the case for our
\ac{BBH} systems. This is due to the fact that the higher-mass ratios available to NSBH systems allow
for cases where the black-hole spin angular momentum is larger than the orbital angular momentum.
For the values of this angle that the \ac{BBH} parameter space \emph{can} produce, we see
similar behavior between the \ac{BBH} and NSBH parameter spaces. The main difference is that it is more
likely for signals to have larger values of the angle between $\mathbf{J}$ and $\mathbf{L}$, and
therefore be more likely to show precessional effects, in the NSBH parameter space than in the \ac{BBH}
parameter space.

\begin{figure*}[tb]
  \includegraphics[width=\columnwidth]{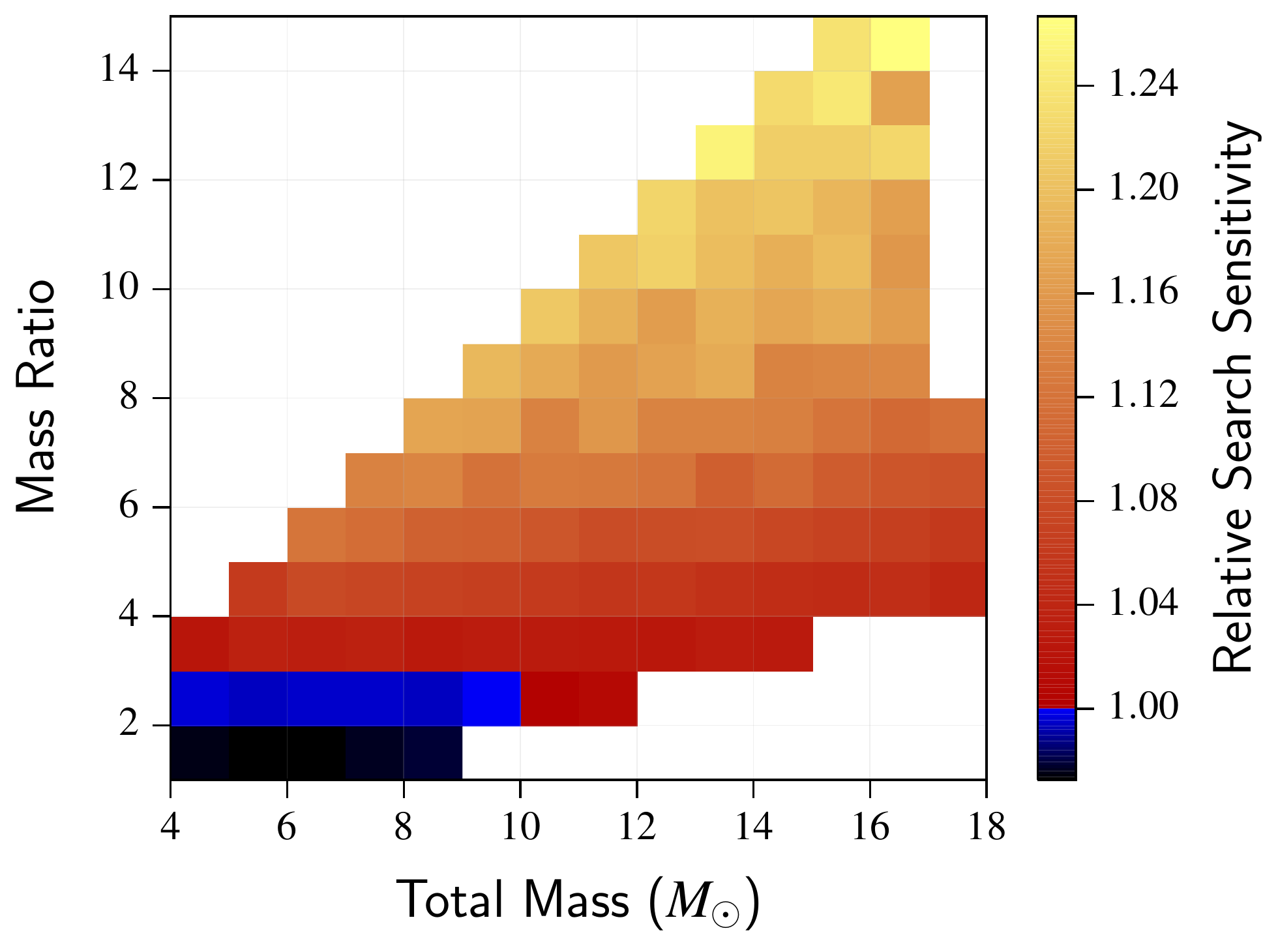}
  \includegraphics[width=\columnwidth]{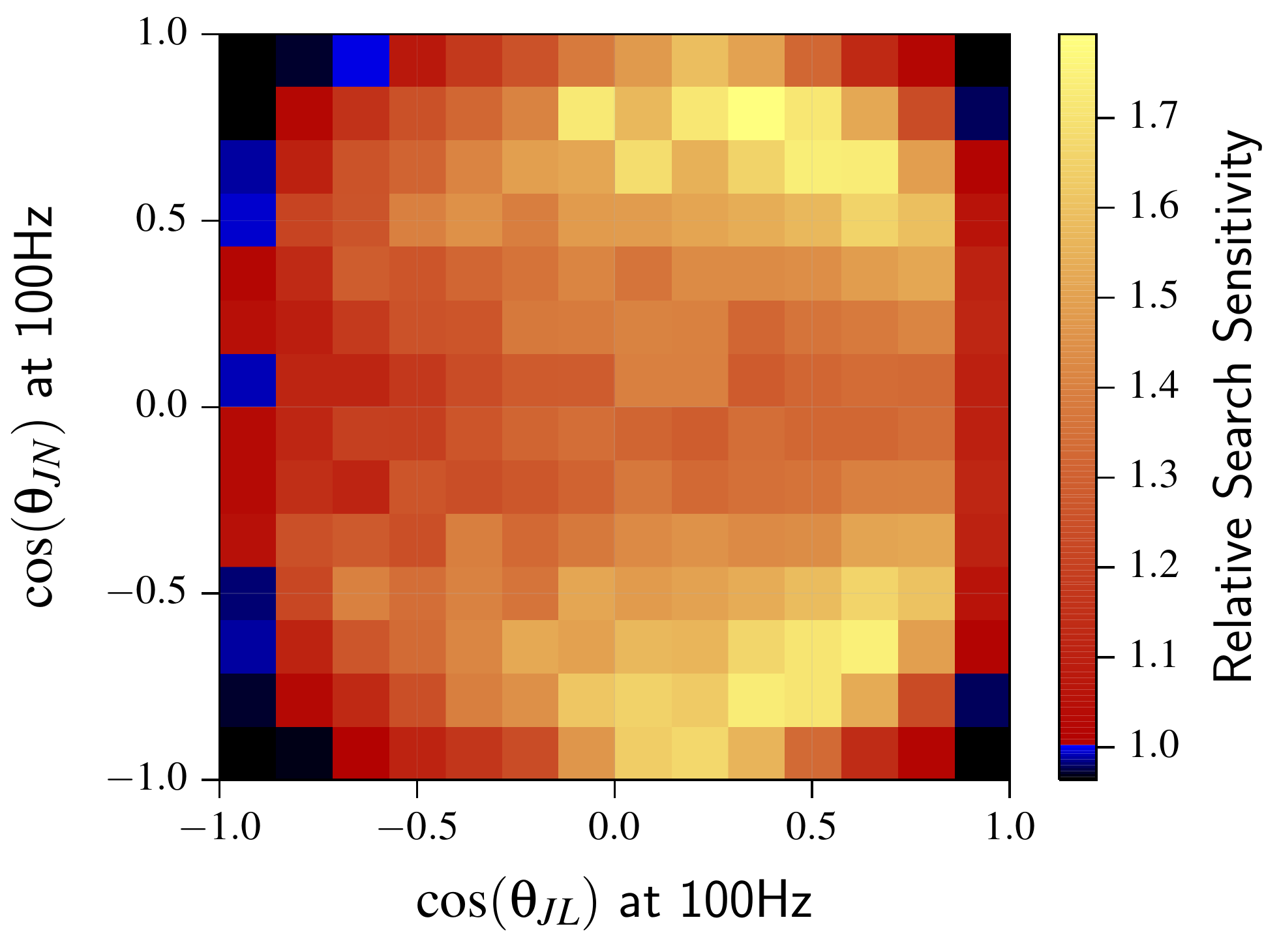}
  \caption{\label{fig:earlynsbh_analysis_results3}This figure is equivalent to
  Fig.~\ref{fig:earlynsbh_analysis_results2} except
  here we assume that for the precessing search \emph{all} fitting factors are unity. This provides us with
  an upper limit on the sensitivity increase that could be gained if the template bank was placed much
  more densely. This is an upper limit because the increased background from the hypothetical denser precessing bank is
  not taken into account.}
\end{figure*}

We can also ask if the fitting factor that we chose to place the precessing bank with is affecting our
results, and if results would improve if the template bank were denser. To try to answer this we reproduce
Fig.~\ref{fig:earlynsbh_analysis_results2}, except when constructing it we set all fitting factors
for the precessing bank to unity. We use the same background increase as given in
Table~\ref{tbl:bkg_thresholds}, which will be an underestimate, but this can provide an upper limit
on the relative search sensitivity that would be obtained by using a denser template bank.
The results of this can be seen in Fig.~\ref{fig:earlynsbh_analysis_results3}. Here the features are
qualitatively similar to those in Fig.~\ref{fig:earlynsbh_analysis_results2}, but the relative sensitivity
improves. 

\section{Discussion}
\label{sec:discussion}

In this work we have (i) derived a new method for detecting compact-binary coalescences when
using waveform filters with generically oriented spins, (ii) constructed
banks of generic-spin template waveforms using this method, (iii) demonstrated the method 
on a stretch of Gaussian noise and (iv) assessed the relative search sensitivity between our new
generic-spin search and the current aligned-spin search. In general,
averaged over the parameter spaces we consider, we have found that using our generic-spin search
does not result in a net increase in the number of detections of compact-binary
mergers---in our NSBH parameter space we saw an increase of only 1\%. However, we have
demonstrated that in regions of parameter space where precessional effects are large, we
can see improvements in detection rate that are larger than 50\%. Systems where the precessional
effects are strong may be rare, but these are also systems which offer us a better chance to disentangle the
various parameters that describe the source, which in turn could allow us a better chance
to understand the nature and origin of these systems~\cite{Vitale:2014mka,O'Shaughnessy:2014dka}. Therefore, one might
argue that an observation of a highly precessing system might be worth more than the one,
or several, observations of systems that do not exhibit precessional effects.

We have also demonstrated that when considering systems with GW150914-like masses and
generically oriented spins, we find signal recovery fractions that are larger than
0.95. This is consistent with what is expected due to the minimal match of 0.97 that is
used to set the discreteness of the template bank in the aligned-spin parameter space.
Current Advanced LIGO searches are therefore not missing systems with masses similar to GW150914
because the waveform filters do not consider misaligned spins.

It is foreseeable that in future work alternative precessing search methods
might be proposed that could improve on the formulation we provide here.
However, there is a fundamental difficulty we have observed
in this work that leads us to believe that it will not be possible to
significantly improve the relative sensitivity of such a hypothetical search over the one described here.
Specifically, we notice that the number
of templates needed to adequately cover a precessing parameter space is at
least an order of magnitude larger than that required to cover the aligned-spin parameter space.
This happens even though we are using a considerably smaller fitting factor for our precessing
template bank than the aligned-spin one. From this, one concludes that the \emph{size}
of the precessing parameter space is significantly larger than the region of that
parameter space that is covered by a template bank placed in only the aligned-spin manifold.
However, the majority of our signal waveforms, which assume an isotropic distribution of the
spin directions, are recovered well by aligned-spin template waveforms. This implies that
the density of astrophysical systems that lie close to the aligned-spin region of parameter
space is also significantly larger than the density of astrophysical systems in the remaining,
large, region of parameter space where precession is important. For these reasons, it
is difficult to gain a significant increase in the number of signals observed when covering
the significantly larger, sparsely populated, precessing parameter space unless the distribution
of signals in the Universe strongly prefers highly precessing cases.

In this work, we have not discussed signal-based consistency tests. Real \ac{GW}
data are not Gaussian, and non-Gaussianities often produce a high \ac{SNR} in matched-filter outputs.
This will be equally true for aligned-spin as for precessing template banks. To be able
to separate real events from non-Gaussianities with large SNRs we use
a set of signal-based consistency tests~\cite{Babak:2012zx, Allen:2004gu}. Ranking
statistics are then constructed that combine the \ac{SNR} with the signal-based
consistency tests such that non-Gaussianities are down-weighted if they do not
match the features we expect of real signals~\cite{Babak:2012zx}.
To be able to apply our methods on real data,
we need to extend these signal-based consistency tests to our
precessing search. This will be our focus in future work.

It is worth pointing
out that the signal-based consistency tests down-weight both non-Gaussianities \emph{and}
real signals that have a low overlap with the best matching template waveform~\cite{Allen:2004gu}.
It is therefore possible that
our results will improve slightly when signal-based consistency tests are included. However,
there is also the possibility that the noise background will increase for precessing
templates relative to aligned-spin templates when applied in real noise even with signal-based
consistency tests. This would decrease the improvement of using the precessing template bank.

Our results are also dependent on how well the models we are using match the waveforms
that will be produced by real compact-binary mergers. For our \ac{BBH} parameter space we are using a waveform model
including inspiral, merger and ringdown phases and which has been tuned against NR 
simulations~\cite{Hannam:2013oca,Khan:2015jqa}. For our NSBH space, we are using an
inspiral-only model, which is known to have some disagreement with other
inspiral-only models, with inspiral-merger ringdown models and with NR~\cite{Nitz:2013mxa, Kumar:2015tha}.
In cases where there is mismatch between our models and the real signal the signal recovery fractions
will be lower than what we have calculated here, but this will be true for both aligned-spin and
precessing template waveforms. Nevertheless the methods presented here are equally applicable
as waveform approximants improve.

For both our \ac{BBH} and NSBH waveform models we have restricted them to only consider the dominant
$l=2, |m|=2$ modes of the \ac{GW} in the coprecessing frame. The effect of
using higher-order modes for nonprecessing waveforms in searches has been previously
explored~\cite{Capano:2013raa}, but no computationally feasible search method has been proposed.
It would be possible to extend the method described here by including also the initial phase
of the binaries in the orbital plane as a discrete template bank parameter. This would alter our
\ac{SMS}, but would enable us to filter with precessing, higher-order mode
waveforms. Exploring how computationally costly this would be and whether this method
would be feasible at all is a topic we leave for future work.

For both the \ac{BBH} and NSBH parameter spaces we consider here we find an order of magnitude increase
between the number of templates in the aligned-spin bank and the number of templates in the
precessing bank. Coupled with the fact that filtering precessing templates using our scheme is
a factor of 2 more expensive than filtering aligned-spin templates results in searches using
our precessing banks being a factor of 20 more computationally expensive than searches using
aligned-spin template banks. This additional computational cost could be used to, for example,
increase the fitting factor of the aligned-spin bank and potentially also gain some increase
in search sensitivity~\cite{Keppel:2013yia}. We would again make the point that the observation of a
highly precessing signal could be very astrophysically rewarding, but exploring how to reduce
this computational cost would be very beneficial. One possible approach we wish to consider
in the future is, when building the template bank, to not include templates corresponding to
systems with a low intrinsic luminosity. This might allow us to reduce the template count \emph{and}
the noise background. Other possibilities include some form of hierarchical approach to filtering
the templates in the template bank. One could also consider schemes to make the bank construction
process more efficient, such as by parallelizing not only in chirp mass bins, but also
by using bins in mass ratio~\cite{HANNOVERPREC}.

\section*{Acknowledgments}

The authors would like to thank Alex Nitz and Lijing Shao for useful and 
detailed comments, and Stas Babak, Sebastian Khan, Harald Pfeiffer, Michael Pürrer, Vivien Raymond, B. Sathyaprakash,
Patricia Schmidt and Andrea Taracchini for helpful discussions. The authors thank the
anonymous referees for detailed and thoughtful comments on the manuscript. The authors would also like to thank
the Max Planck Gesellschaft for support.

\appendix

\section{Computational optimizations of precessing bank construction}
\label{sec:app_compopt}

To generate our stochastic template banks of precessing signals, we use a
number of recent optimizations described in Refs.~\cite{Ajith:2012mn,Fehrmann:2014cpa,
Capano:2016dsf}, as well as some new methods. We briefly describe these optimizations
in the next paragraphs.

When choosing $h_\mathrm{prop}$ the masses are chosen from a 
uniform distribution in 
the chirp time coordinates $\tau_0$-$\tau_3$~\cite{Sathyaprakash:1991mt}. These coordinates are
optimal for nonspinning, inspiral-only signals. They are
suboptimal in our case, but better than choosing mass parameters
uniformly in $m_1$-$m_2$, since typically many more templates are
needed at low mass compared to high mass.
Spins are chosen isotropically
with a uniform distribution of spin magnitude.
Sky locations and orientation parameters are chosen
isotropically. These distributions may not be the optimal choices
for stochastic precessing template bank construction, but we
leave a further exploration of this for future work.

When computing matches, those templates in the template bank
that have values of chirp mass closest to the proposed template
have matches computed first. This allows one to reject proposed templates
more quickly. In addition a match is only computed if the difference in chirp mass
between the template in the bank and the point being tested is
less than $25\%$. This can lead to some overcoverage, especially at
higher masses.

When computing matches, a match is first computed using a
frequency spacing in the overlap integral of $df=8$Hz.
If the resulting overlap is 0.05 less
than the desired minimal match, the match at smaller values of the
frequency spacing
is assumed to also be less than the desired minimal match and is
not computed explicitly.
If the match using a frequency spacing of 8Hz is not less than
the minimal match minus 0.05 then the match is computed using frequency
spacings of $df=0.5$Hz and $df=0.25$Hz.
If these matches agree to within 0.1\% then the $df=0.25$Hz value is used.
Otherwise an additional match
at $df=0.125$Hz is computed and compared against $df=0.25$Hz. This process
continues until either the match is in agreement to
0.1\% or the value of $df$ has exceeded the
inverse of the waveform length. The first test using $df=8$Hz quickly rejects
points that do not match well to each other. The subsequent tests at
much smaller values of $df$, $0.5$Hz and $0.25$Hz, are because precessional
features in the waveform can be missed when using larger values of $df$ and 
we found some cases where matches would agree well at $df=4$Hz and $df=2$Hz, but
then diverge when $df$ became smaller.

A coarse bank is first produced with a lower convergence
threshold to roughly map out the density of templates. Here, we
choose a coarse threshold such that the algorithm terminates when
fewer than one in $X$ proposed templates are accepted to the bank
over the last ten iterations. The space is then split into $Y$
equally spaced bands in chirp mass, and a process launched for
each chirp mass band, choosing points only in that band, and filling
templates until fewer than one in $Z$ proposed templates are accepted to the bank
over the last ten iterations. It is possible that
the parallel processes will cover the same region, but this effect is
mitigated by beginning the process with a partially complete template bank
over the full space before parallelizing. The parallelization also
helps to ensure that all parts of the parameter space are being
sampled by the stochastic process. The values $X$, $Y$ and $Z$ are
chosen empirically and vary for the different banks generated here.

Further optimizations of this process will be greatly desirable if using
the methods described in this paper in the future. We leave that to
future work.

\bibliography{references}

\end{document}